\documentclass[final,5p,twocolumn]{elsarticle}
\usepackage[binary-units=true]{siunitx}
\newcommand{\SIUS}[2]{\SI[separate-uncertainty=true]{#1}{#2}}
\usepackage{lineno}
\modulolinenumbers[5]
\usepackage{amsmath,amssymb,amsfonts}
\usepackage{xfrac}
\usepackage{algorithmic}
\usepackage{graphicx}
\usepackage{textcomp}
\usepackage[table]{xcolor}
\usepackage{booktabs} 
\usepackage{multirow}

  \usepackage[sfdefault,light]{FiraSans} 
  \usepackage[T1]{fontenc}

\definecolor{tableGray}{HTML}{E5E5E5}
\definecolor{col1}{HTML}{1E88E5}
\definecolor{col2}{HTML}{D81B60}
\definecolor{col3}{HTML}{43A047}
\definecolor{col4}{HTML}{F4511E}

\usepackage[normalem]{ulem}
\usepackage{xspace}
\newcommand{\ie}{i.e.,\xspace}
\newcommand{\eg}{e.g.,\xspace}

\newcommand{\arm}{Arm\xspace}
\newcommand{\armv}{Armv8\xspace}
\newcommand{\mn}{MareNostrum4\xspace}
\newcommand{\dibona}{Dibona\xspace}
\newcommand{\txtwo}{ThunderX2\xspace}
\newcommand{\skylake}{Skylake\xspace}

\newcommand{\dibonaarm}{Dibona-TX2\xspace}
\newcommand{\dibonaintel}{Dibona-X86\xspace}
\newcommand{\dbnarm}{DBN-TX2\xspace}
\newcommand{\dbnintel}{DBN-X86\xspace}

\newcommand{\graph}{Graph 500\xspace}

\usepackage[colorlinks=true, allcolors=blue]{hyperref}

\usepackage[utf8]{inputenc}

\RequirePackage[normalem]{ulem}
\RequirePackage{color}\definecolor{RED}{rgb}{1,0,0}\definecolor{BLUE}{rgb}{0,0,1}

\providecommand{\DIFdel}[1]{{\protect\color{RED}\sout{#1}}}
\providecommand{\DIFdel}[1]{}

\journal{
Future Generation Computer Systems (FGCS)
}

\begin{document}

\begin{frontmatter}

\title{
{\bf Performance and energy consumption of HPC workloads\\
on a cluster based on \arm \txtwo CPU }
}


\author[bsc]{Filippo Mantovani\corref{mycorrespondingauthor}}
\ead{filippo.mantovani@bsc.es}
\cortext[mycorrespondingauthor]{Corresponding author}

\author[bsc]{Marta Garcia-Gasulla}
\ead{marta.garcia@bsc.es}

\author[hlrs]{Jos\'e Gracia}
\ead{gracia@hlrs.de}

\author[unican]{Esteban Stafford}
\ead{esteban.stafford@unican.es}

\author[bsc]{Fabio Banchelli}
\ead{fabio.banchelli@bsc.es}

\author[bsc]{Marc Josep-Fabrego}
\ead{marc.josep@bsc.es}

\author[bsc]{Joel Criado-Ledesma}
\ead{joel.criado@bsc.es}

\author[hlrs]{Mathias Nachtmann}

\address[bsc]{Barcelona Supercomputing Center (Spain)}
\address[hlrs]{High Performance Computing Center Stuttgart (HLRS), 
  University of Stuttgart (Germany)}
\address[unican]{University of Cantabria (Spain)}

\begin{abstract}
In this paper, we analyze the performance and energy consumption of an \arm-based high-performance computing (HPC)  system developed within the European project Mont-Blanc~3.
This system, called \dibona, has been integrated by ATOS/Bull, and it is powered by the latest Marvell's CPU, \txtwo. This CPU is the same one that powers the Astra supercomputer, the first \arm-based supercomputer entering the Top500 in November 2018.
We study from micro-benchmarks up to large production codes. We include an interdisciplinary evaluation of three scientific applications (a finite-element fluid dynamics code, a smoothed particle hydrodynamics code, and a lattice Boltzmann code) and the \graph benchmark, focusing on parallel and energy efficiency as well as studying their scalability up to thousands of \armv cores. 
For comparison, we run the same tests on state-of-the-art x86 nodes included in \dibona and the Tier-0 supercomputer \mn.
Our experiments show that the \txtwo has a 25\% lower performance on average, mainly due to its small vector unit yet somewhat compensated by its 30\% wider links between the CPU and the main memory.
We found that the software ecosystem of the \armv architecture is comparable to the one available for Intel.
Our results also show that \txtwo delivers similar or better energy-to-solution and scalability, proving that \arm-based chips are legitimate contenders in the market of next-generation HPC systems.

\end{abstract}


\end{frontmatter}


\begin{minipage}[t]{\textwidth}
\vspace{-15.5cm}
\centering
\includegraphics[width=0.1\linewidth]{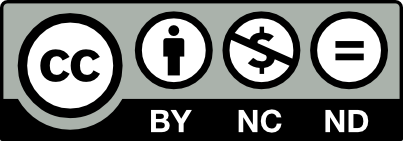}\\
\footnotesize{©2020. This manuscript version is made available under the CC-BY-NC-ND 4.0 license.\\
\url{http://creativecommons.org/licenses/by-nc-nd/4.0/}\\
\url{https://doi.org/10.1016/j.future.2020.06.033}}
\end{minipage}

\vspace{-1cm}

\section{Introduction}\label{secIntro}


The \arm architecture is gaining significant traction in the race to Exascale. 
Several international collaborations, including the Japanese Post-K, the European Mont-Blanc, and the UK's GW4/EPSRC, announced the adoption of \arm technology as a viable option for high-end production high-performance computing (HPC)  systems.
During November 2018, for the first time, an \arm-based system was ranked in the Top500 list. It was the Astra supercomputer powered by Marvell’s (former Cavium) \txtwo processor, integrated by HPE and installed at the Sandia National Laboratories (US).
For more than six years, research projects have evaluated \arm-based systems for parallel and scientific computing in collaboration with industry, advocating the higher efficiency of this technology mutated from the mobile and the embedded market. 

Computational requirements of scientific and industrial applications are increasing. As a consequence, the HPC market is also growing steadily in double digits, according to one of the latest reports of Hyperion Research\footnote{\url{
https://www.hpcwire.com/2018/06/27/at-isc18-hyperion-reports
}}.
This market growth goes hand in hand with the appearance of a jungle of new technologies and architectures. From an economic standpoint, this will help to diversify the market and level the prices. From a technical point of view, however, it is not always clear what   the gains are, in terms of performance and energy consumption, of new technologies entering the data centers. The most prominent example of this phenomenon is the adoption of GP-GPUs in HPC: even if the benefit of using graphical accelerators was shown in the early 2000, it is only now, after more than ten years, that data centers consider GP-GPU a well established HPC technology. A very similar dynamic is happening with the adoption of \arm CPUs. 

In this blurry scenario, we performed our study to help HPC application scientists in understanding the real implications of using this new technology. We targeted the latest \arm-powered CPU by Marvell that offers close to state-of-the-art performance. The \txtwo processor powers the \dibona cluster, the system we have used for our evaluation. In contrast, we have also considered Intel \skylake CPUs, which are available on some nodes of \dibona, as well as in the state-of-the-art Tier-0 supercomputer \mn.
Our analysis follows a bottom-up approach: we start from the micro-benchmarking of the cluster, moving then to a higher level evaluation using HPC applications, first in a single node, and finally scaling up to a thousand cores. We selected a set of production HPC workloads from those considered in the European project Mont-Blanc 3, which proved to be scalable on different state-of-the-art HPC architectures, and we measured their performance at scale as well as their energy footprint. We also include the \graph benchmark as a representative of an emerging irregular work flow.
By doing so, we de facto explore different architectural configurations of CPUs and   show that, while the ISA does not seem to make a big difference in the final overall performance, the micro-architectural choices (\eg the size of the SIMD units or the organization of the memory hierarchy) and the software configurations (\eg compilers) are vital factors for delivering a powerful and efficient modern HPC system. 



The main contributions of this paper are
\begin{enumerate}[i]
  \item We provide a comparative analysis of the Arm-based \txtwo cluster \dibona and the Intel \skylake based Tier-0 supercomputer \mn. We show that the software tools for the \txtwo are mature and on the same level as those available for \skylake, and the performance of a \txtwo node is, on average, 25\% less than for a \skylake node.
  \item We analyze the power drain of the Marvell \txtwo processor under three HPC production codes and a benchmark, comparing it with that of state-of-the-art HPC technology. We show that the energy-to-solution of the two cluster technology under study are equivalent.
  \item We study the behavior at scale of these codes in a \txtwo cluster and x86-based Tier-0 supercomputer \mn, concluding that the scalability trends of both technologies are comparable.
\end{enumerate}


The rest of the document is structured as follows:
In Section~\ref{secRelatedWork}, we introduce the context of our evaluation, and 
in Section~\ref{secHwIntro} and~\ref{secHw}, we detail the evaluation methodology and the hardware features of the HPC clusters used in the following sections.
Section~\ref{secEval} is dedicated to introducing the HPC applications and their characterization when running on \dibona.
Section~\ref{secEnergy} is reserved for energy measurements and comparisons.
In Section~\ref{secScalability}, we report the result of our tests at scale on \dibona and \mn Tier-0 HPC supercomputer,
and Section~\ref{secProjections} includes performance projections based on these results.
We conclude with Section~\ref{secConclusions}, where we summarize our evaluation experience.

\section{Related Work}\label{secRelatedWork}

Several papers have been published with preliminary analyses of benchmarks and performance projections of the \arm-based system-on-chip (SoC) coming from the mobile and embedded market 
\cite{
rajovic2014tibidabo,
rajovic2013supercomputing,
rajovic2016mont,
oyarzun2018efficient,
cesini2018infn
}.
More recently, tests on the \arm-based server SoC have also appeared in the literature~\cite{
calore2018advanced,
puzovic2016quantifying
}.
%
%
The most relevant and recent work focusing on evaluating the \txtwo processor is the one of McIntosh-Smith et al. in~\cite{mcintosh2018performance}. They evaluate Isambard, a high-end Tier-2 system developed by Cray in the framework of the GW4 alliance~\cite{gw4}. While they provide an extensive single-node evaluation, we complement their contribution evaluating HPC applications at scale.

For our evaluation, we chose three HPC production codes:
{\em Alya}~\cite{vazquez2016alya}, a finite elements code handling multi-physics simulations developed at the Barcelona Supercomputing Center, already studied on different architectures in~\cite{garcia2018computational,garcia2019runtime};
a Lattice-Boltzmann code {\em LBC}~\cite{gracia2012lbctasks} using the BGK approximation for the collision term~\cite{bhatnagar1954model}, which, even if it is not a full-production code, mimics the typical behavior of Lattice Boltzmann simulations used both for advanced fluid dynamics studies~\cite{
biferale2011second
}
and architectural evaluation~\cite{
mantovani2013exploiting,
calore2019optimization
};
{\em Tangaroa}~\cite{Tangaroa,reinhardt2017asyncSPH}, a smoothed particle hydrodynamics code, whose purpose is to simulate fluids in a way suitable for computer animation.
We also included in our study the {\em \graph} benchmark~\cite{murphy2010introducing} as a representative of an emerging irregular workflow that   not only benefits from the pure floating-point performance of a system, but also stresses the memory and the network with irregular access patterns. In addition to the several studies of this benchmark, we recall the one of Checconi et~al. in~\cite{checconi2013massive}, where the authors study the \graph benchmark on a successful HPC architecture, the IBM Blue Gene/Q CPU.

The use of \arm-technology in HPC has also been moved by energy-efficient arguments. Like several others, we try to address them comparing the performance and the energy-to-solution of our four test cases on different architectures.
Radulovic et al. \cite{radulovic2018mainstream} provide an extensive study of emerging HPC architectures, including their performance and efficiency. However, they focus mostly on benchmarks and kernels, while we evaluate complex codes used for the production of scientific results (\eg Alya).
Jarus et al. \cite{jarus2013performance} also study the performance and efficiency of CPUs for HPC by different manufacturers. They focused on HPL, providing for each CPU an extrapolation of the ranking in the Green500. As already mentioned, our contribution focuses less on benchmarks and more on real applications. It is also worth mentioning that the CPU technology that we are evaluating is targeted at the data center market, not a technology borrowed from the embedded world.
D'Agostino et al. \cite{d2019soc} as well as McIntosh-Smith et al. \cite{mcintosh2018performance} also analyze the cost efficiency of emerging technologies in HPC. We prefer to leave the variable of the price out of our analysis, since it involves a negotiation process beyond our control and, in our opinion, would make the comparison less relevant.

\section{Hardware Description}\label{secHwIntro}

In this section, we present the technical specifications of the systems used for our analysis. These are the new HPC \arm-based cluster, called \dibona, mainly built around the \txtwo processor, and for comparison, \mn,  an Intel-based supercomputer, which represents a well-known baseline HPC architecture. 
Table~\ref{tabHwClusterSummary} shows a summary of the hardware configurations of the clusters used as a reference in our study.

\begin{table}[htbp]
\caption{Summary of the hardware configuration of the platforms}
\resizebox{\columnwidth}{!}{%
\rowcolors{2}{gray!15}{white}
\begin{tabular}{l|c|c|c}
                     & \textbf{\dibonaarm} & \textbf{\dibonaintel} & \textbf{\mn} \\ \midrule
Core architecture    & Armv8               & Intel x86           & Intel x86             \\
CPU name             & \txtwo              & \skylake Platinum   & \skylake Platinum     \\
Frequency {[}GHz{]}  & 2.0                 & 2.1                 & 2.1                   \\
Sockets/node         & 2                   & 2                   & 2                     \\
Core/node            & 64                  & 56                  & 48                    \\
Memory/node {[}GB{]} & 256                 & 256                 & 96                    \\
Memory tech.         & DDR4-2666           & DDR4-3200           & DDR4-3200             \\
Memory channels      & 8                   & 6                   & 6                     \\
Num. of nodes        & 40                  & 3                   & 3456                  \\
Interconnection      & Infiniband EDR      & Infiniband EDR      & Intel OmniPath        \\
System integrator    & ATOS/Bull           & ATOS/Bull           & Lenovo               
\end{tabular}%
}
\label{tabHwClusterSummary}
\end{table}


\subsection{The \mn Supercomputer}\label{secMN4}

\mn is a Tier-0 supercomputer in production at Barcelona Supercomputing Center (BSC) in Barcelona, Spain. It has a total of 3456 compute nodes available, which include two Intel Xeon Platinum 8160 processors, each with 24~\skylake cores clocked at 2.1~GHz and with 6~DDR4-3200 memory channels~\footnote{\url{https://www.bsc.es/user-support/mn4.php}}. Each node is equipped with 96~GB of DDR4-3200. The interconnection network is 100 Gbit/s Intel Omni-Path (OPA). \mn runs a Linux~4.4.12 kernel, and it uses SLURM~17.11.7 as a workload manager. In this supercomputer, we perform our scalability study presented in Section~\ref{secScalability}.


\subsection{The \dibona Cluster}\label{secDibona}

The \dibona cluster is the primary outcome of the European project Mont-Blanc 3. It was designed and integrated by ATOS/Bull and evaluated by the Mont-Blanc 3 partners between September 2018 and February 2019~\cite{mb3DeliverableAppicationsDibona}. Thanks to the ATOS/Bull Sequana HPC infrastructure, it seamlessly integrates 40 \arm-based compute nodes, together with three x86-based compute nodes.
Each \arm-based compute node is powered by two Marvell \txtwo CN9980 processors, each with 32~\armv cores at 2.0~GHz, 32~MB L3 cache, and 8 DDR4-2666 memory channels. The x86 nodes are powered by Intel Xeon Platinum 8176 processor with 28~\skylake cores running at 2.1~GHz with 6~DDR4-2666 memory channels. The total amount of RAM installed is 256~GB per compute node.

Compute nodes are interconnected with a fat-tree network, with a pruning factor of \sfrac{1}{2} at level 1, implemented with Mellanox IB EDR-100 switches.
A separated 1~GbE network is employed for the management of the cluster and a network file system (NFS).
\dibona runs the Linux~4.14.0 kernel, and it uses SLURM~17.02.11 patched by ATOS/Bull as a job scheduler.


\subsection{Mapping of our study on hardware resources}

Having \mn deployed at the Barcelona Supercomputing Center allows us to use it as a reference machine to be compared with \dibona for benchmarking (as described in Sections~\ref{secHw} and~\ref{secEval}) and as a platform to scale applications at thousands of cores (as presented in Section~\ref{secScalability}).

Moreover, the fact that \dibona houses \armv and x86 compute nodes allowed us to carry out fair energy comparisons in Section~\ref{secEnergy} with nodes on the same system, differing mainly in their CPU architecture, and offering an identical power monitoring infrastructure.

In Table~\ref{tabSummaryTable}, we map the studies performed in this manuscript onto the hardware platforms available for this evaluation.

\begin{table}[htbp]
\caption{Summary of the study performed in this manuscript}
\resizebox{\columnwidth}{!}{%
\begin{tabular}{l|l|c|c|c}
\textbf{Type of study}                                                                                                      & \textbf{Application} & \textbf{\dibonaarm} & \textbf{\dibonaintel} & \textbf{\mn} \\ \midrule
\multirow{4}{*}{\begin{tabular}[c]{@{}l@{}}Hardware\\ characterization\\ Section~\ref{secHw}\end{tabular}}                  & STREAM               & \checkmark                   &                     & \checkmark                \\
                                                                                                                            & FPU $\mu$-kernel     & \checkmark                   &                     & \checkmark                \\
                                                                                                                            & Roofline             & \checkmark                   &                     & \checkmark                \\
                                                                                                                            & OSU                  & \checkmark                   &                     & \checkmark                \\ \midrule
\multirow{4}{*}{\begin{tabular}[c]{@{}l@{}}Single-node\\ performance\\ Section~\ref{secEval}\end{tabular}}                  & Alya                 & \checkmark                   &                     & \checkmark                \\
                                                                                                                            & LBC                  & \checkmark                   &                     & \checkmark                \\
                                                                                                                            & Tangaroa             & \checkmark                   &                     & \checkmark                \\
                                                                                                                            & \graph               & \checkmark                   &                     & \checkmark                \\ \midrule
\multirow{4}{*}{\begin{tabular}[c]{@{}l@{}}Energy\\ measurements\\ Section~\ref{secEnergy}\end{tabular}}                    & Alya                 & \checkmark                   & \checkmark          &                           \\
                                                                                                                            & LBC                  & \checkmark                   & \checkmark          &                           \\
                                                                                                                            & Tangaroa             & \checkmark                   & \checkmark          &                           \\
                                                                                                                            & \graph               & \checkmark                   & \checkmark          &                           \\ \midrule
\multirow{4}{*}{\begin{tabular}[c]{@{}l@{}}Scalability\\ Section~\ref{secScalability}\end{tabular}}                         & Alya                 & \checkmark                   &                     & \checkmark                \\
                                                                                                                            & LBC                  & \checkmark                   &                     & \checkmark                \\
                                                                                                                            & Tangaroa             & \checkmark                   &                     & \checkmark                \\
                                                                                                                            & \graph               & \checkmark                   &                     & \checkmark               
\end{tabular}%
}
\label{tabSummaryTable}
\end{table}

In the rest of the document, we refer to the partition of Dibona powered by Arm \txtwo CPUs with the name {\em \dibonaarm} (shortened to \dbnarm when required).
Artworks related to this systems are displayed in red.
The partition of Dibona powered by x86 \skylake CPUs is named {\em \dibonaintel} (shortened to \dbnintel when required). 
Artworks related to x86-based CPUs (both \dibonaintel and \mn) are displayed in blue.

\section{Hardware Characterization}\label{secHw}
This section is dedicated to the micro-benchmarking \dibonaarm, the \arm-based platform selected for our study. We use \mn as the state-of-the-art system for comparison.


\subsection{\dibonaarm Memory Subsystem}\label{secMemory}

Here we evaluate the memory bandwidth using STREAM~\cite{mccalpin1995memory}, a simple synthetic benchmark to measure sustainable memory bandwidth. Our study analyses a \txtwo node from the \dibonaarm cluster, compared side by side with a \skylake node from \mn. STREAM kernels iterate through data arrays of double-precision floating-point elements (8~bytes) with a size fixed at compile time. 
The size of each array $E$ must be greater than 
the maximum between 
ten million elements and
four times the size of the sum of all the last-level caches.
$$ E \ge max~\left\{  10,000,000 ~;~ 4 \cdot S / 8 \right\} $$
where $E$ is the number of elements of each array, and $S$ is the size of the last level cache in bytes.
Table~\ref{tab:mem_brief} shows a brief overview of the memory subsystem of each socket, including the minimum value of $E$ and the compiler flavor and version used. The reader should remember that each node has two sockets.

\begin{table}[htbp]
\centering
\caption{Memory subsystem overview}
\resizebox{\columnwidth}{!}{%
\rowcolors{2}{gray!15}{white}
\begin{tabular}{ l | r | r }
                & {\bf \dibonaarm}  & {\bf \mn} \\ \midrule
Architecture    & Arm \txtwo    & x86 \skylake       \\ 
L1 cache size   & 32 kB         & 64 kB          \\
L2 cache size   & 256 kB        & 256 kB         \\
L3 cache size   & 32 MB         & 33 MB          \\
Main mem. tech. & DDR4-2666     & DDR4-3200      \\
\# of channels  & 8             & 6              \\
Peak bandwidth  & 170.64 GB/s   & 153.60 GB/s    \\
$E$             & 16777216      & 17301504       \\
Compiler        & Arm 18.3      & Intel 17.0.4   \\
\end{tabular}
}
\label{tab:mem_brief}
\end{table}

\begin{figure}[htbp]
  \centering
  \includegraphics[width=\columnwidth]{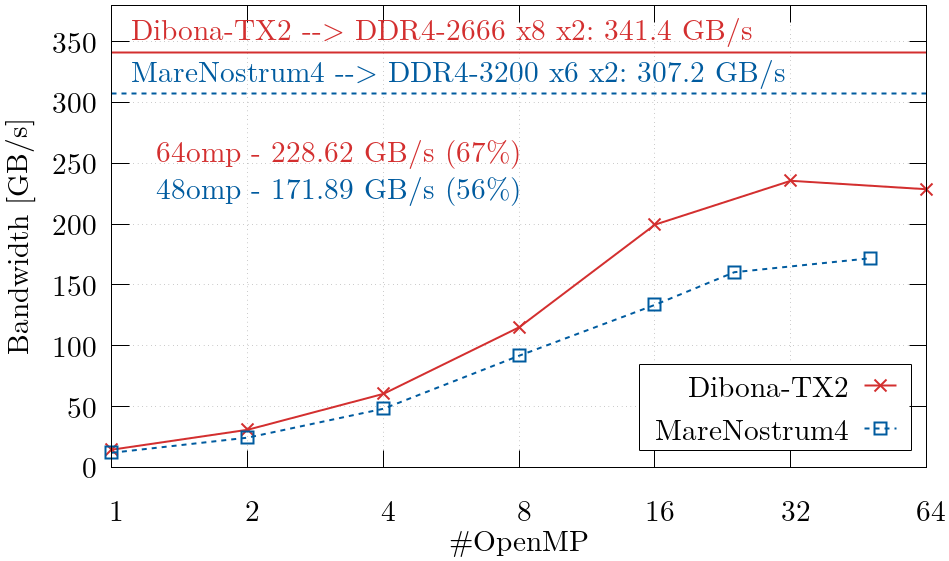}
  \caption{STREAM Triad Best bandwidth achieved over number of OpenMP threads in \dibona and \mn nodes, thread binding: Interleaved}
  \label{fig:stream_bwXomp_interleaved}
\end{figure}

We run the benchmark by fixing the problem size to the minimum valid value of $E$ for each platform and increasing the number of OpenMP threads.
We report the results of the {\em Triad} function as a representative kernel since the rest of them have a similar behavior.
Threads are pinned to cores by using \verb|OMP_PROC_BIND=true|, distributing the threads evenly across both sockets and minimizing the number of threads accessing the same L2 cache. Figure~\ref{fig:stream_bwXomp_interleaved} shows the achieved bandwidth.
The $x-axis$ represents the number of OpenMP threads, growing up to the number of cores in each node, and the $y-axis$ indicates the maximum bandwidth achieved throughout 200 executions of the kernel.
The figure also includes two horizontal lines representing the theoretical peak bandwidth of each processor. Please note that the DDR technology is different: the \txtwo CPU housed in the \dibonaarm cluster uses DDR4-2666, with a theoretical peak of 21.33~GB/s per channel, and the \skylake CPU of \mn uses DDR4-3200, with a theoretical peak of 25.60~GB/s per channel.

The figure shows that \txtwo reaches 228.62~GB/s (67\% of the peak) in the Triad kernel when running with 64~OpenMP threads (\ie one full node), while \mn obtains 171.89~GB/s (56\% of the peak) with 48~OpenMP threads. This fact shows that, although the \txtwo has slower memory, it outperforms the \skylake thanks to the extra memory channels. 


\subsection{Floating-Point Throughput}\label{secFloat}

We designed a micro-kernel to measure the peak floating-point throughput of a given CPU. We call this code {\em FPU\_$\mu$kernel}. It contains exclusively fused-multiply-accumulate assembly instructions with no data dependencies between them. The kernel has four versions distinguishing between {\em i)} scalar and vector instructions  and {\em ii)} single and double precision.
The \armv ISA has floating-point instructions that accept single- and double-precision registers as operands. In this case, the kernel uses the instruction \verb|FMADD|. In \dibonaarm, the \armv cores of the \txtwo integrate the NEON vector extension. With 128~bit registers, it can fit either two double-precision or four single-precision data elements per register. The kernel uses the NEON vector instruction \verb|FMLA|.
Although the x86 ISA has floating-point instructions that run on the FPU,  the more recent SIMD instructions of the AVX512 vector extension are recommended. Thus, the compiler automatically translates \verb|a * b + c| to \verb|VFMADD132SS| or \verb|VFMADD132SD| for single- and double-precision instructions, respectively. We implemented the SIMD version of the kernel using the AVX512 instruction \verb|VFMADD132PS| for single precision and the \verb|VFMADD132PD| for double precision. This means that the scalar version of the code in the x86 architecture will use vector instructions with the same behavior as scalar floating-point instructions.
The theoretical peak of the vector unit can be computed as the product of {\em i)} the vector size in elements (\eg four single-precision elements in NEON); {\em ii)}~the number of instructions issued per cycle; {\em iii)}~the frequency of the processor; {\em iv)}~the number of floating-point operations made by the instruction (\eg fused-multiply-accumulate does two floating-point operations).
\begin{table}[htbp]
\centering
\caption{Theoretical peak performance of one NEON and one AVX512 vector units in \dibonaarm and \mn nodes}
\resizebox{\columnwidth}{!}{%
\rowcolors{2}{gray!15}{white}
\begin{tabular}{ l | r r | r r }
                 & \multicolumn{2}{c|}{\bf \dibonaarm} & \multicolumn{2}{c}{\bf \mn} \\ \midrule
Architecture     & \multicolumn{2}{c|}{Arm \txtwo} & \multicolumn{2}{c}{x86 \skylake} \\ 
Vector extension & \multicolumn{2}{c|}{NEON} & \multicolumn{2}{c}{AVX512} \\ 
Instruction      & \multicolumn{2}{c|}{FMLA} & \multicolumn{2}{c}{VFMADD} \\ 
Precision        & Single     & Double     & Single            & Double   \\
Vec. Length      & 4          & 2          & 16                & 8        \\
Issue/cycle      & 2          & 2          & 2                 & 2        \\
Freq. [GHz]      & 2.00       & 2.00       & 2.10              & 2.10     \\
Flop/Inst        & 2          & 2          & 2                 & 2        \\
Peak [GFlop/s]   & 32.00      & 16.00      & 134.40            & 67.20    \\
\end{tabular}
}
\label{tab:vec_limit}
\end{table}
Table~\ref{tab:vec_limit} lists these parameters and the theoretical peak for the vector extensions available in \dibonaarm and \mn nodes, both with single- and double-precision vector operations.

\begin{figure}[htbp]
  \centering
  \includegraphics[width=\columnwidth]{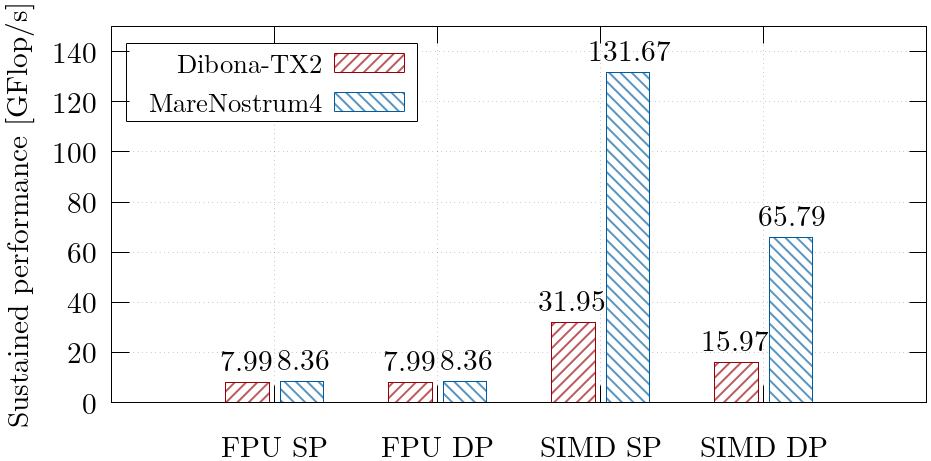}
  \caption{Sustained performance in one core of the four versions of the FPU\_$\mu$Kernel in \dibonaarm and \mn nodes (see theoretical peak performance in Table~\ref{tab:vec_limit})}
  \label{fig:fpu}
\end{figure}

Figure~\ref{fig:fpu} shows the results measured on both machines. It can be seen that the floating-point units of both architectures have similar performance, and that the slightly better performance of \mn is solely due to its 5\% higher CPU frequency. In contrast, the AVX512 vector unit of \skylake housed in \mn outperforms the NEON unit of the \txtwo processor in \dibonaarm by a factor of $\times4.22$ in single precision and $\times4.11$ in double precision. The fact that the vector registers of the \skylake are four times larger than those of the \txtwo, together with the increased frequency, account for this performance difference. 


\subsection{Roofline Model}\label{secRoofline}

The roofline model allows us to visualize the hardware limitations of different computational systems and characterize the workload of computational kernels~\cite{ofenbeck2014applying}.
The model plots the sustained performance $P_s(I)$, measured in (Giga) floating-point operations per second, as a function of the arithmetic intensity $I$.
The arithmetic intensity $I$ is defined as $ I = \frac{W}{D} $, where $W$ is the number of floating-point operations executed by the application (\ie computational workload), while $D$ is the number of bytes that the application exchanges with the main memory (\ie the application dataset). 

\begin{equation}
P_s(I) = %
\begin{cases}
F_p         \quad \quad \quad \,\, \text{if}\,\, I > I_r \\
B_p \cdot I \quad \quad            \text{if}\,\, I \le I_r 
\end{cases}
\label{eqRoofline}
\end{equation}

\begin{figure}[htbp]
\centering
  \centering
  \includegraphics[width=\linewidth]{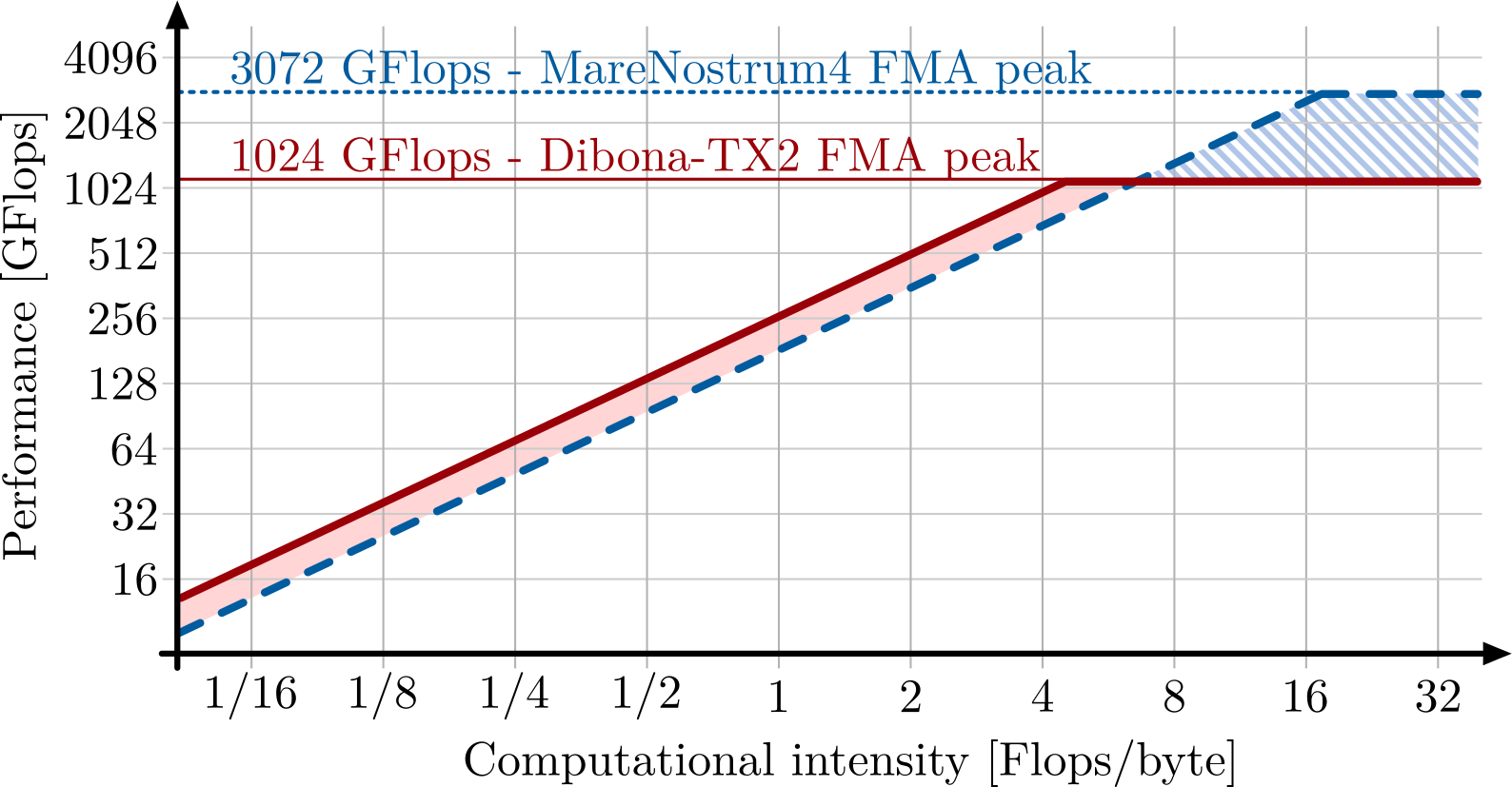}
  \caption{Roofline of \dibonaarm (red) and \mn (blue)}
  \label{figRoofline}
\end{figure}

Equation~\ref{eqRoofline} shows the theoretical formulation of the roofline model.
$F_p$ and $B_p$ are, respectively, the peak floating-point performance (expressed in Flop/s) and the peak memory bandwidth to/from memory (expressed in Byte/s), while $I_r = F_p / B_p$.
Using $F_p$ as measured with the FP-$\mu$kernel in Section~\ref{secFloat} and $B_p$ as benchmarked in Section~\ref{secMemory}, we can, therefore, compute $P_s(I)$ and plot it in Figure~\ref{figRoofline}, where we present the roofline model of a \dibonaarm node and a \mn node.

An engineer knowing the arithmetic intensity and the computational performance of its application or kernel can, therefore, identify the space available for its optimization towards the maximum performance delivered by the system.

Figure~\ref{figRoofline} quickly reveals the different architectural configurations of the two types of node  and allows the exploration of a slightly different configuration space. The red area is indeed a configuration space enabled by the 8-channel memory sub-system of the \txtwo in \dibonaarm, not reachable with current \skylake architectures that are only offering six memory channels. In contrast, the area highlighted in blue is the space where highly vectorizable codes can take advantage of the wider SIMD units on the \skylake (AVX512) compared to the narrower \armv NEON extension. 


\subsection{Interconnection Network}\label{secNetwork}

\begin{figure}[htbp]
  \centering
  \includegraphics[width=\linewidth]{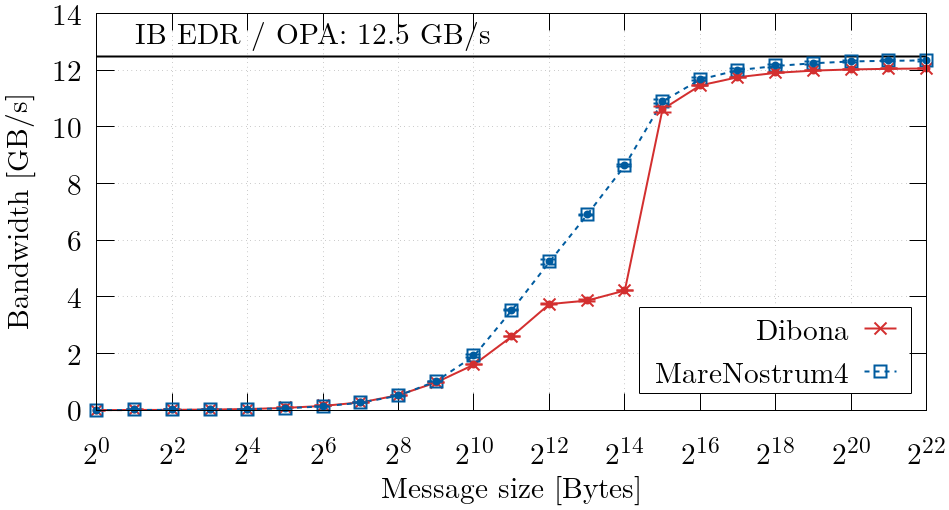}
  \caption{Bandwidth between two processes in different nodes}
  \label{figOsuBw}
\end{figure}

To this point, we have characterized different architectural aspects of the \txtwo and \skylake processors, but multi-node scaling experiments will be affected by the interconnection network. Therefore, this section evaluates the network performance of the \dibonaarm cluster, which uses Mellanox Infiniband EDR (IB) interconnect. Naturally, we compare it with that of the \mn supercomputer, which uses an Intel~Omni-Path (OPA) interconnect.
Figure~\ref{figOsuBw} shows the achieved throughput, as reported by the OSU benchmarks~\cite{liu2003performance} as a function of the message size of the communication. All points represent the average value of $100$ repetitions of the communication.

Both networks approach the theoretical peak as the message size increases ($\sim$95\%). It seems that OPA is consistently achieving a better bandwidth than IB with message sizes over 256~KiB. The difference in bandwidth is also very noticeable at message sizes around 4~KiB and 8~KiB, where OPA almost doubles IB. 
It seems that the measured bandwidth of OSU in \dibonaarm stalls around~8 and~16~KiB but then shoots up to 10~GB/s for larger message sizes. This behavior is consistent throughout multiple pairs of nodes and between executions. 
We verified that this behavior disappears if we measure the bandwidth with the \verb!ib_read_bw! tool by Mellanox. As this tool exchanges data using the raw network protocol, we can only conclude that the ``valley'' appearing in Figure~\ref{figOsuBw} is caused by the OpenMPI configuration deployed by ATOS/Bull on the Dibona cluster at the moment of the tests.

\begin{figure}[htbp]
\centering
  \includegraphics[width=\columnwidth]{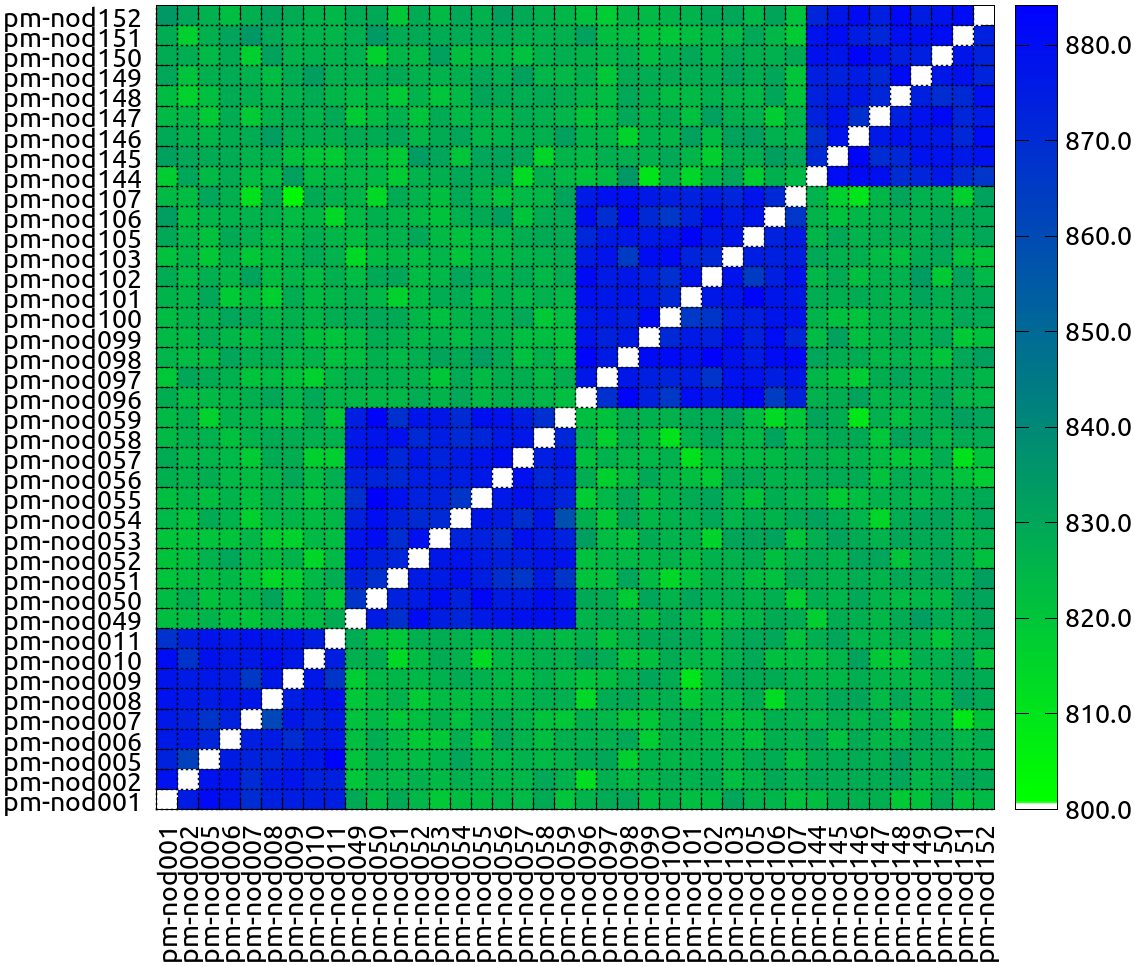}
\caption{Weak links in the \dibonaarm network, message size: 4~KiB, axis correspond to nodes, the bandwidth [MB/s] in color code}
\label{fig:weak_links_4k}
\end{figure}

We repeated the tests for multiple pairs of nodes to determine if there were systematic weak links on \dibonaarm. Figure~\ref{fig:weak_links_4k} shows a map where the axes represent the pair of nodes, and each cell is color-coded to indicate the bandwidth. We present the measurements for message sizes of 4~KiB. There is a recurring pattern along the diagonal where pairs of nodes have higher bandwidth. This fact is due to the network topology, as these pairs of nodes are connected to the same level-1 switch. Nodes that are topologically farther apart reach 10~\% less bandwidth than neighboring nodes.

\section{Application Characterization}\label{secEval}

In this section, we describe the three scientific applications and the benchmark we used to evaluate \dibonaarm. 
It provides a short description and a computational characterization of the codes, together with a performance evaluation at a small scale (one or two compute nodes).
The applications were selected from the set of production HPC workloads that were analyzed in the European project Mont-Blanc~3. They proved to be scalable on different state-of-the-art HPC architectures and represent a range of real scientific HPC applications. The lattice Boltzmann code (LBC)  operates on a stencil-like access pattern,  Tangaroa is a particle tracking code with a behavior similar to sparse-matrix applications, and Alya is a complex finite-element code that comprises several solvers with different characteristics. All of them are parallelized leveraging MPI using two-sided primitives.

We also evaluate the \graph benchmark with the goal of including an evaluation of an irregular code representative of emerging HPC workloads that are becoming increasingly important for diverse fields, such as social networks, biology, intelligence, and e-commerce.
We invite the reader interested in other benchmark evaluations to complement our work with \eg that of McIntosh-Smith et al. \cite{mcintosh2018performance} and D.~Ruiz et al. \cite{HPCS19_hpcg}. We also highlight the work of G.~Ramirez-Gargallo et al. \cite{HPML19_tensorflow} for the reader interested in the performance of \dibonaarm under artificial intelligence workloads.

For each code, we present a brief description of its purpose, internal structure, and the chosen dataset. 
We include a small scale evaluation of the applications and benchmark on \dibonaarm and \mn nodes. 
It also details the performance of different compiler solutions available in both clusters: the generic GNU suite and the vendor-specific alternatives (LLVM-based \arm HPC Compiler and the Intel suite).
Although, in several cases, the complexity of the codes does not allow a fine-grained benchmarking, we try to provide quantitative observations of computational features to help understanding the behavior of the applications at scale.

Unless otherwise noted, we use two metrics to evaluate the scalability, the elapsed time, which gives the idea of the fastest option, and the efficiency, which helps to understand how well   the code scale does on a given system. In this section, the efficiency $E$ has been computed as follows: $E = \frac{t_1}{t_i \cdot i}$, where $t_1$ is the execution time when running with one core, and $t_i$ is the execution time when running with $i$ cores.

\subsection{Alya}\label{secAlya}

Alya~\cite{vazquez2016alya} is a high-performance computational mechanics code developed at the Barcelona Supercomputing Center. Alya can solve different physics, including incompressible/compressible turbulent flows, solid mechanics, chemistry, particle transport, heat transfer, and electrical propagation. 
It is part of the Unified European Applications Benchmark Suite (UEABS) of PRACE, a set of twelve relevant codes together with their data sets, which can realistically be run on large systems. Thus Alya complies with the highest standards in HPC.

\begin{figure}[htbp]
  \centering
  \includegraphics[width=\linewidth]{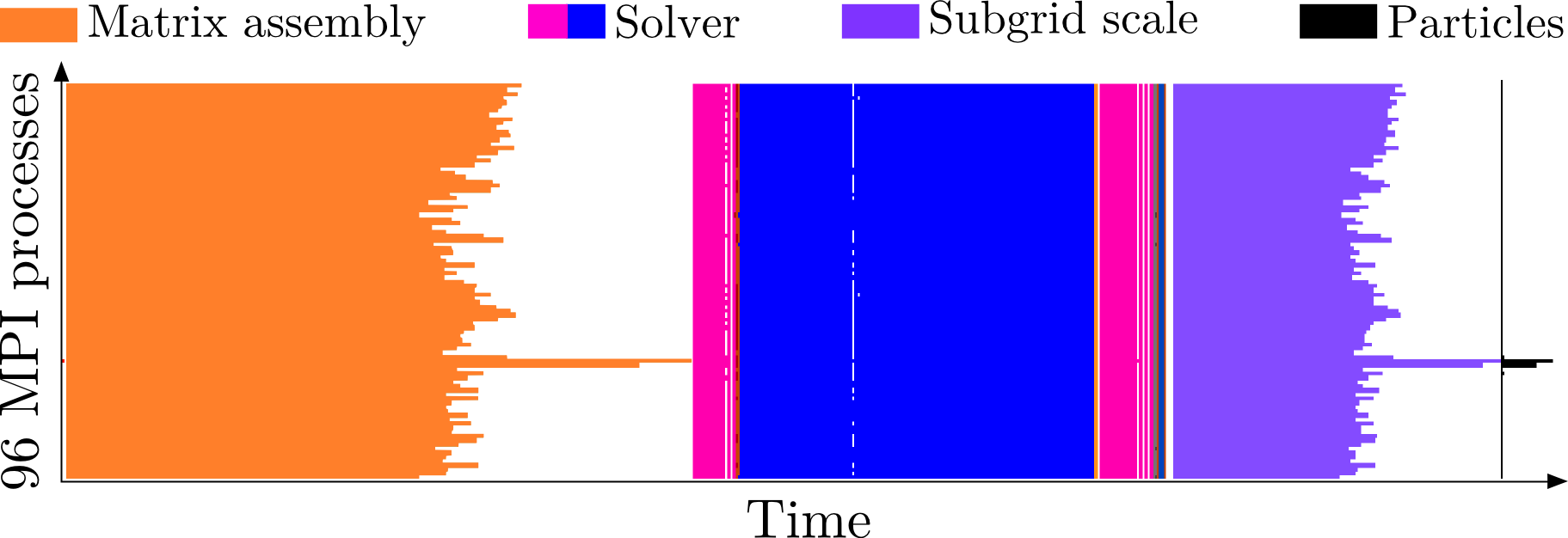}
  \caption{Alya -- Trace with highlight of the main computational phases}
  \label{figAlyaTrace}
\end{figure}

In this work, we simulate an incompressible turbulent flow and the transport of Lagrangian particles with Alya. In particular, we simulate the air through the human respiratory system and the transport of the particles during rapid inhalation~\cite{calmet2016large}.

In Figure~\ref{figAlyaTrace}, we can see a trace of a one-time step of the respiratory simulation with Alya. The time is represented on the $x$-axis, and on the $y$-axis are the different MPI processes. The color indicates the phase that is being executed, and white corresponds to MPI communication. The matrix assembly, algebraic solver, and subgrid scale phases correspond to the computation of the fluid (the velocity of the air), and the particles phase corresponds to the calculation of the transport of particles. 

In the trace, we highlight  the most time-consuming phases. In Table~\ref{tabAlyaArithIntensity}, we quantify the percentage of the total time spent in each one.

\begin{table}[htbp]
\centering
\caption{Alya -- Percentage of the total execution time and arithmetic intensity for different phases of the respiratory simulation executed with 96~MPI processes}
\rowcolors{2}{gray!15}{white}
 \begin{tabular}{l|r|r}
\textbf{Phase}  & \textbf{\% Time} & {\bf I} \\\midrule
Matrix assembly &          40.84\% &    0.09 \\
Solver1         &          16.13\% &    0.03 \\
Solver2         &           4.20\% &    0.12 \\
Subgrid scale   &          21.43\% &    0.07 \\
Particles       &           3.37\% &    0.05
\end{tabular}
 \label{tabAlyaArithIntensity}
\end{table}

The rightmost column of Table~\ref{tabAlyaArithIntensity} shows the \emph{arithmetic intensity} $I$ for each phase (see Section \ref{secRoofline}). 
We measured the computational load $W$ as the number of double-precision operations reported by the CPU counters. We also determined the data set $D = (L + S) \cdot 8$, where $L$ and $S$ are the number of load and store instructions accessing double-precision data values (8 Bytes) measured by the hardware counters of the CPU.
We show in Table~\ref{tabAlyaArithIntensity} the values of $I$, arithmetic intensities, of the different phases of Alya, ranging from one floating-point operation per 8 bytes transferred in the first type of solver to one floating-point operation per 32 bytes transferred in the second solver of Alya. 

\subsubsection{Experimental setup}

We performed a small scale comparison of Alya executing on one and two nodes of \dibonaarm and \mn.
The idea is to provide a small-scale evaluation and compare at the same time hardware platforms and compiler solutions, as summarized in Table~\ref{tabAlyaSoftwareEnvironments}.
\begin{table}[htbp]
\centering
\caption{Alya -- Software environment used on different clusters}
\rowcolors{2}{gray!15}{white}
\begin{tabular}{l|c|c}
{\bf Platform } & {\bf Compiler} & {\bf MPI version}  \\ \midrule
\dibonaarm  & GNU 7.2.1      & OpenMPI 2.0.2.11  \\
\dibonaarm  & \arm 18.4.2    & OpenMPI 2.0.2.14  \\
\mn         & GNU 7.2.1      & OpenMPI 2.0.2.10  \\
\mn         & Intel 2018.1   & OpenMPI 2.0.2.10
\end{tabular}
\label{tabAlyaSoftwareEnvironments}
\end{table}

Production simulations can run for up to $10^5$ time steps. The results presented in this evaluation have been obtained by averaging 10 time steps, like the one shown in Figure~\ref{figAlyaTrace}. Statistical variability of the measurements is below 1\%, so we chose not to pollute plots with error bars.

The mesh used in our experiments is a subject specific geometry including from the face to the seventh branch generation of the bronchopulmonary tree. The mesh is hybrid with 17.7 million elements, including prisms, tetrahedra, and pyramids. We partition the mesh with METIS~\cite{parMETIS} to distribute the elements as homogeneously as possible across different MPI processes. We inject 400,000 particles during the first time step of the simulation through the nasal orifice.

\subsubsection{Node-to-node comparison}\label{secAlyaNodeToNodeComp}

\begin{figure}[htbp]
  \centering
  \includegraphics[width=\linewidth]{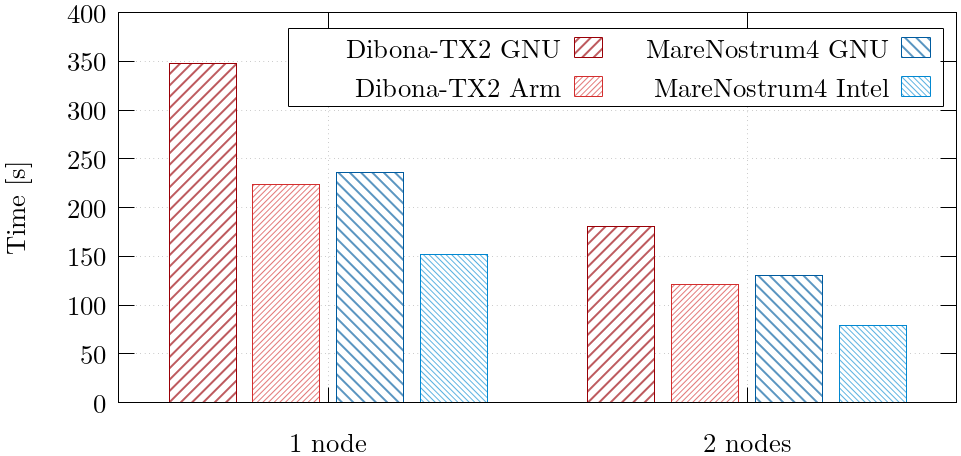}
  \caption{Alya -- Performance running on one and two nodes of \dibonaarm and \mn}
  \label{figAlyaPerf}
\end{figure}

In Figure~\ref{figAlyaPerf}, we report the average elapsed time of a time step when running with different compilers on both node types of the \dibonaarm and \mn.
Comparing both platforms with each compiler family, it can be seen that \mn shows a 30\% improvement over \dibonaarm regardless of the compiler.
Furthermore, for each platform, vendor-specific compilers deliver better performance: a 34\% average improvement on \armv \txtwo, and 37\% on x86 \skylake. 
In all cases, the efficiency when going from one to two nodes is between 90\% and 96\%.

\subsection{LBC}\label{secLBC}

LBC is a Lattice Boltzmann code written in Fortran used for advanced fluid dynamics studies, using the BGK approximation for the collision term. 
The version used in this work is pure-MPI with two-sided communication only. We
have benchmarked the code on a single node of \dibonaarm and \mn \skylake.
Multi-node scalability will be presented in a later section.

\subsubsection{Experimental setup}

The following details regarding the experimental setup are common to LBC
experiments and performance data reported below.
On both architectures, we used the GNU suite to compile the code and
OpenMPI for communication across nodes. Besides, we used compilers by
the processor vendor, \ie Arm and Intel, respectively, to evaluate the impact of
the compiler on the performance of the code. Details of the software stack can
be found in Table~\ref{tabLBCSoftwareEnvironments}.

\begin{table}[htbp]
\centering
\caption{LBC -- Software environment used on clusters}
\rowcolors{2}{gray!15}{white}
\begin{tabular}{l|c|c}
{\bf Platform } & {\bf Compiler} & {\bf MPI version}  \\ \midrule
\dibonaarm  & GNU 7.2.1      & OpenMPI 2.0.2.14  \\
\dibonaarm  & \arm 19.1      & OpenMPI 2.0.2.14  \\
\mn         & GNU 7.2.0      & OpenMPI 3.1.1  \\
\mn         & Intel 2018.1   & OpenMPI 3.1.1
\end{tabular}
\label{tabLBCSoftwareEnvironments}
\end{table}

Each data point in this section's figures corresponds to the average of over at
least 30 time measurements. To get a representative result, at most 10
runs were done within the same job, and jobs were distributed over two or more
days.  We   also calculated the standard deviation of the sample as error
estimates. However, we do not plot error bars, as they are usually small and
would only crowd the figures. In cases were errors are substantial, we mention
this fact in the text.

In the Lattice Boltzmann community, the underlying grid cells are often referred
to as lattice elements. The usual metric for performance is the \emph{number
of lattice updates per time interval} measured in units of \emph{MLUP/s} (mega
lattice updates per second, $10^6$ lattice updates per second). This metric is
reported by the application at the end of the run. Note that LBC disregards the
initialization phase and other overhead when reporting performance.

For comparison purposes, the problem size of all experiments is chosen such that
the number of lattice elements per core present on the node is $n_C = (256 \times 256 \times 32)$.
The problem sizes assigned to each node is $n_N = n_C \cdot C$, where $C$ is the number of cores per node.
We call $p_N$ the 3-dimensional domain decomposition of $C$ that represents the MPI ranks assigned to each node.
The memory requirement is roughly proportional to the total number of lattice elements (including one ghost cell
at each boundary); each lattice elements stores 41 double-precision float numbers.
Table~\ref{tab:LBC_1node} shows the domain configurations for each of the cluster.

\begin{table*}[tbph]
\centering
\caption{Node-to-node comparison for LBC on \dibonaarm and \mn nodes}
\rowcolors{2}{gray!15}{white}
  \begin{tabular}{l|r|r|r|r}
                                  & \multicolumn{2}{c|}{\bf \dibonaarm}                     & \multicolumn{2}{c}{\bf \mn}    \\ \midrule
  Cores per node, $C$             & \multicolumn{2}{c|}{ 64 }                               & \multicolumn{2}{c}{48}          \\
  Domain size per node, $n_N$     & \multicolumn{2}{c|}{ $512\!\times\!512\!\times\!512$ }  & \multicolumn{2}{c}{ $512\!\times\!512\!\times\!384$ } \\
  Domain decomposition, $p_N$     & \multicolumn{2}{c|}{ $2 \!\times\! 2 \!\times\! 16$ }   & \multicolumn{2}{c}{ $2 \!\times\! 2 \!\times\! 12$  } \\
  Memory requirements             & \multicolumn{2}{c|}{ \SI{41}{\gibi\byte} }              & \multicolumn{2}{c}{ \SI{31}{\gibi\byte}  }            \\
  Compiler                        & GNU                   & Arm                             & GNU                   & Intel           \\
  Performance \si{[MLUP/\second]} & \SIUS{265.4+-3.0}{} & \SIUS{196.1+-1.0}{}               & \SIUS{277.7+-1.4}{}   & \SIUS{279.8+-1.4}{}        \\
  \end{tabular}
\label{tab:LBC_1node}
\end{table*}

The code was run for 10 time steps without any intermediate output. This proved
sufficiently large for accurate time measurements.

\subsubsection{Node-to-node comparison} 

To compare the computing capabilities of the \txtwo and \skylake processors, we executed the pure-MPI version of LBC on single nodes of \dibonaarm and \mn.

\begin{figure}[htbp]
  \includegraphics[width=\columnwidth]{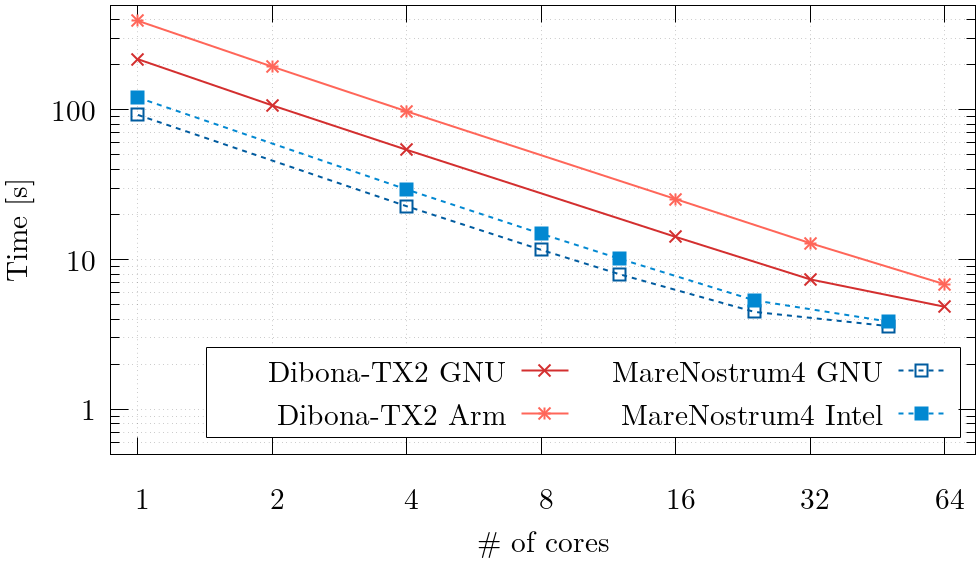}
  \caption{LBC -- Strong scaling on \dibonaarm and \mn, with execution time as a function of number of cores}
  \label{fig:lbc_dibona_strong_scaling_time}
\end{figure}

\begin{figure}[htbp]
  \includegraphics[width=\columnwidth]{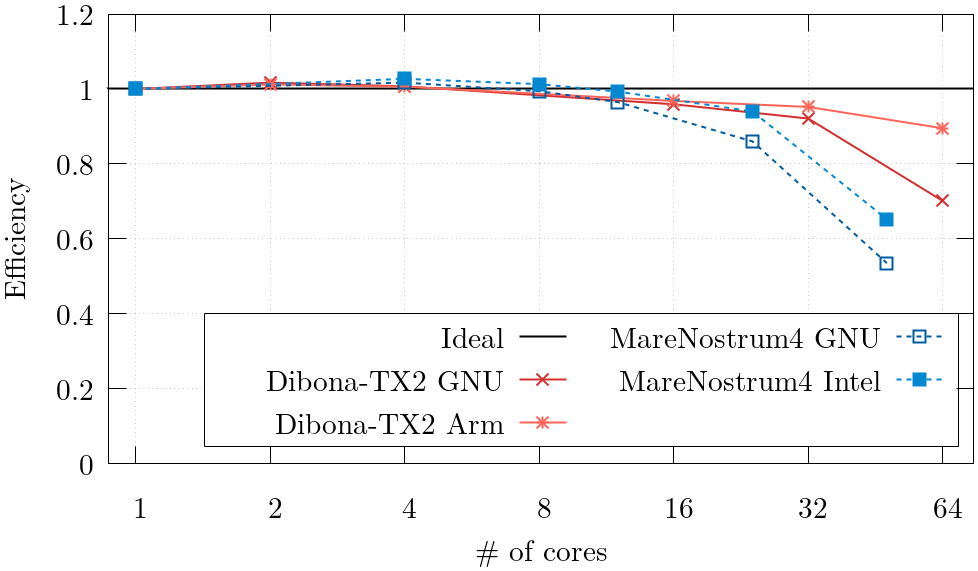}
  \caption{LBC -- Strong scaling on \dibonaarm and \mn, with parallel efficiency as a function of number of cores}
  \label{fig:lbc_dibona_strong_scaling_eff}
\end{figure}

We assume that the MPI communication within a single
node will have little impact on the performance of the code. Single-core runs
were disregarded, as one core is not sufficient to saturate the available memory
bandwidth on either system, and such performance measurements would overestimate
and obfuscate the real application performance in production runs.

The performance of single-node runs is reported in Table~\ref{tab:LBC_1node} in
terms of the application-specific metric.
In addition, we performed a strong scaling experiment on a single node of \dibonaarm and \mn. 
MPI ranks have been placed to spread evenly across NUMA domains to best utilize the available memory bandwidth.
The execution time is illustrated in Figure~\ref{fig:lbc_dibona_strong_scaling_time}, while the efficiency is shown in Figure~\ref{fig:lbc_dibona_strong_scaling_eff}.

With the GNU compiler, the performance
on the \dibonaarm is 4\% lower than on the \mn node, even though the
number of cores per compute node is roughly \SI{33}{\percent} higher.
This indicates that, proportionally, \skylake cores outperform \txtwo cores.

We also note that the performance of the code heavily depends on the maturity of the compiler technology used.
On \dibonaarm, the performance with the \arm compiler is roughly \SI{25}{\percent} lower than the performance obtained with the GNU compiler (see Table~\ref{tab:LBC_1node}).
We can only explain this performance degradation with the maturity of the Arm compiler technology since we have seen that the vendor-specific compiler can improve the performance of other applications (see \eg Alya in Section~\ref{secAlyaNodeToNodeComp}).
On \mn, the performance with the GNU compiler and with the Intel compiler is practically identical.

All versions of the code show an initial super-scalar behavior up to roughly 8
cores. This is because   the memory size of the working set per core decreases
  with an increasing core count so cache reuse increases. However,
this effect is countered, as the effective memory bandwidth per core saturates
at larger core counts. 

While both kinds of nodes have a similar core count, the \txtwo architecture on \dibonaarm retains higher
strong scaling efficiency. Again, we see slight differences amongst compilers,
in particular, the Intel compiler produces code that scales better at larger core
counts of \mn nodes than the GNU compiler. 
For the Arm compiler, we observe high efficiencies even up to the full core
count. This, however, is an artifact of the low performance of this particular
version: due to the low compute speed, there is never sufficient pressure on the
memory systems to feel the limited effective bandwidth.

Overall, however, our results suggest that the \txtwo memory system of the \dibonaarm cluster is
relatively better suited for memory bound workloads such as LBC. 

\subsection{Tangaroa}\label{secTangaroa}

Tangaroa is a C++ application that simulates fluid dynamics in a way suitable for computer animation~\cite{Tangaroa,reinhardt2017asyncSPH}. Its results are not meant to be physically correct but only visually plausible. The simulation progresses by analyzing the behavior of a large amount of particles in discrete time steps. The 3D space that contains the particles is partitioned to allow parallel execution of the simulation. Particle position, velocity, and other properties are calculated with single-precision floating-point arithmetic (4 Bytes).

\subsubsection{Experimental setup}

The data extracted from the Tangaroa executions considers the actual simulation of the particles as the region of interest; job setup and data allocation times are not taken into account. Tangaroa tries to hide communication as much as possible; therefore, computation is very intense in this region. Single-node experiments were made on dedicated nodes, meaning that intra-node MPI communication was not perturbed by traffic from other applications. Moreover, since execution times range from tens of minutes to hours, the perturbation from the OS is considered negligible. Each data point shown is the average of five independent executions. In all experiments, the I/O operations were disabled to avoid interference from different network file system technologies and thus improve the accuracy of the measurements. 

The dataset represents a fairly even distribution of around 12 million particles in a box-shaped region. The internal representation of these particles is at least 1.5~GB; there are other data structures whose size is not directly proportional to the number of particles. The size of the dataset is enough to justify using several hundred processes, although a larger one would allow better scalability.

\begin{table}[htbp]
\centering
\caption{Software environment used on clusters for Tangaroa}
\rowcolors{2}{gray!15}{white}
\begin{tabular}{l|c|c}
{\bf Platform } & {\bf Compiler} & {\bf MPI version}  \\ \midrule
\dibonaarm  & GNU 7.2.1      & OpenMPI 3.1.2  \\
\dibonaarm  & \arm 19.1      & OpenMPI 3.1.2  \\
\mn         & GNU 7.2.0      & OpenMPI 3.1.1  \\
\mn         & Intel 2017.4   & IntelMPI 2017.3
\end{tabular}
\label{tabTangaroaSoftwareEnvironments}
\end{table}

Table~\ref{tabTangaroaSoftwareEnvironments} reports the software configuration used for the evaluation of Tangaroa.

\subsubsection{Node-to-node comparison}

The first experiment compares the performance of full nodes, meaning that the same problem is divided into all the cores in each node. The domain containing the particles must be adequately divided between the cores. To accommodate the number of cores in one node of \mn, we had to choose an irregular domain decomposition, which nevertheless led to a reasonably even distribution of particles per core. Experiments with regular domain decompositions did not improve the load balance. 

\begin{table}[h!]
\centering
\caption{Node-to-node comparison for Tangaroa on \dibonaarm and \mn nodes}
\resizebox{\columnwidth}{!}{%
\rowcolors{2}{gray!15}{white}
  \begin{tabular}{l|r|r|r|r}
                       & \multicolumn{2}{c|}{\bf \dibonaarm}              & \multicolumn{2}{c}{\bf \mn}    \\ \midrule
  Cores/node           & \multicolumn{2}{c|}{ 64 }                        & \multicolumn{2}{c}{48}          \\
  Domain decomposition & \multicolumn{2}{c|}{ $4\!\times\!2\!\times\!8$ } & \multicolumn{2}{c}{ $4\!\times\!2\!\times\!6$ }  \\
  Particles/core       &  \multicolumn{2}{c|}{ $ 194240\pm21760 $ }       & \multicolumn{2}{c}{ $ 255664\pm29141 $  } \\
  Compiler             & GNU                & Arm               & GNU                 & Intel           \\
  Simulation time [s]  & 101.16             & 93.77               & 55.84               & 61.20         \\
  IPC                  & 1.64               & 1.66                & 2.62                & 2.47          \\
  \end{tabular}
}
\label{tangaroa_node}
\end{table}

Table \ref{tangaroa_node} summarizes the results of this experiment.
The first two columns of the table show that with the \arm compiler (Clang), the application is 7\% faster than with GCC. Since we observed that the number of SIMD instructions executed by the Clang version is higher than in the GCC one, and that the Instructions per Clock-cycle (IPC) is practically the same, we can conclude that the higher performance of the \arm compiler is due to a better use of the vector units.
With the \skylake node of \mn, however, it is GCC that gives the best performance; through an improvement of the IPC and a higher number of SIMD instructions, it manages an 8\% reduction of the execution time when compared to ICC. 
Comparing both platforms, it can be seen that, although \skylake has roughly 12\% fewer cores than \txtwo, it manages to give a 31\% better performance. This indicates that the computing power of each core in the \txtwo node is substantially less, and that their increased count is not enough to overcome this limitation.

\begin{figure}[htbp]
   \centering
   \includegraphics[width=\columnwidth]{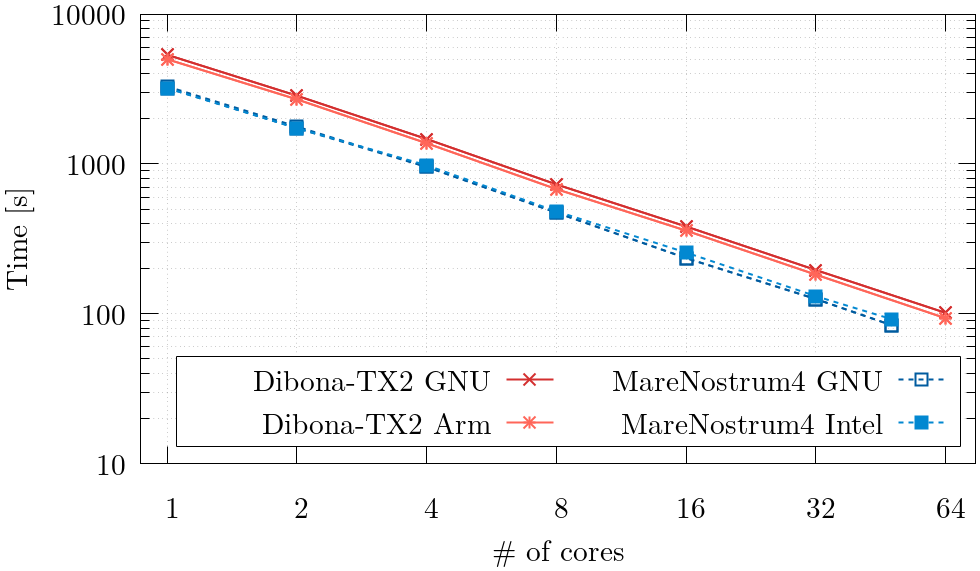}
   \caption{Tangaroa -- Strong scaling on \dibonaarm and \mn, with execution time as a function of number of cores}
   \label{tangaroa_node_time}
\end{figure}

\begin{figure}[htbp]
   \centering
   \includegraphics[width=\columnwidth]{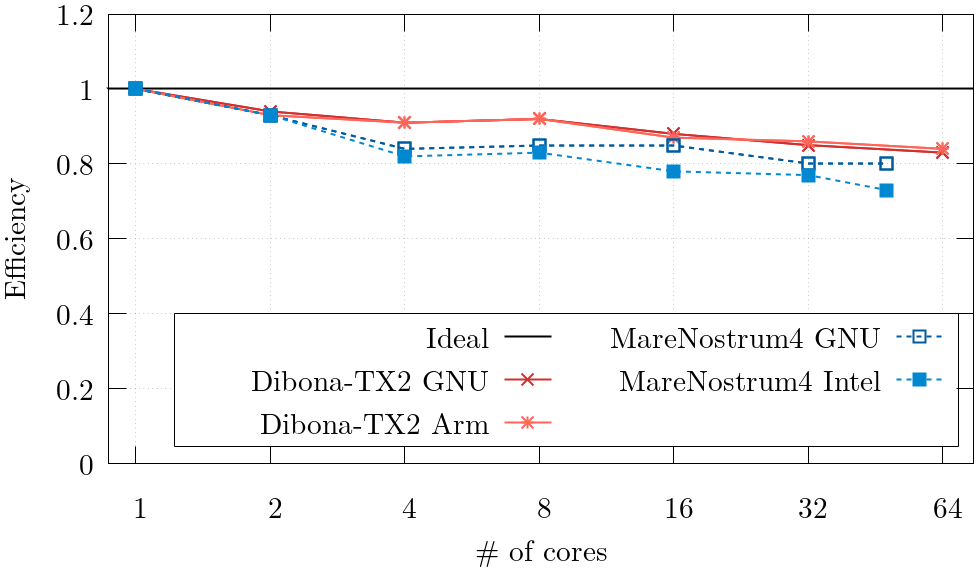}
   \caption{Tangaroa -- Parallel efficiency on \dibonaarm and \mn: the baseline for the efficiency is the time of a single core execution} 
   \label{tangaroa_node_eff}
\end{figure}

The second experiment is a strong scaling test contained within a single node. Process pinning was set to interleaved to maximize memory bandwidth utilization. The results are shown in Figures~\ref{tangaroa_node_time} and~\ref{tangaroa_node_eff}, where it can be seen that both machines diverge from the ideal scaling. This is because each process of Tangaroa must communicate the particles close to the subdomain border to the neighboring processes. Thus, if the number of processes   increases, so does the amount of data that must be transmitted. This has a more noticeable effect on the scalability in \mn node since it has two memory channels less than the \txtwo node.

Comparing the parallel efficiency of Tangaroa in Figure~\ref{tangaroa_node_eff} and the one of LBC in Figure~\ref{fig:lbc_dibona_strong_scaling_eff}, we can also see that the higher pressure to the main memory played by LBC results into a lower parallel efficiency when using all cores of the compute node.

\subsection{\graph}\label{secGraph}

\graph is a benchmark used to rank supercomputers when performing a large-scale graph search problem~\cite{murphy2010introducing}. 
It is written in C, and we evaluated the version parallelized using MPI of the breadth-first search (BFS).
Each graph of the $v$ vertices is associated with a scale $s$, where $s = \log_2 v$. 
The parameter $s$ is the only input parameter that needs to be provided for the benchmark.
The \graph benchmark generates an internal representation of a graph with the number of vertices supplied as input. With this graph, it performs 64 breadth-first searches (BFSs) of randomly generated keys. The numbers presented are the average duration of the 64 BFSs performed and their error, as reported by the benchmark.
While the rules of the benchmark allow one to optimize this internal representation as well as the BFS implementation, we left them ``as is'' in the reference version of the \graph benchmark (v3.0.0).

\subsubsection{Experimental setup}

The \graph benchmark used is optimized to run with a number of MPI processes that is a power of 2. Nevertheless, it offers the possibility of compiling it without this optimization, allowing it to run with a number of processes that is not a power of 2. In our case, the number of cores per node of \mn is not a power of 2; for this reason, we decided to evaluate both options.

In all plots of this section, we report the average time spent in performing 64 BFSs, ignoring the setup and the validation phases of the benchmark.
We selected as input size $s=24$, corresponding to a number of vertices $v = 2^{24}$, which is the largest scale that fits the memory of a single compute node of both clusters and allows a reasonable duration of the simulation.

As for the rest of the applications, we studied the effects of different compilers. 
We summarize the software configurations in Table~\ref{tabGraphSoftwareEnvironments}.

\begin{table}[htbp]
\centering
\caption{\graph\ -- Software environment on different clusters}
\rowcolors{2}{gray!15}{white}
\begin{tabular}{l|c|c}
{\bf Platform } & {\bf Compiler} & {\bf MPI version}  \\ \midrule
\dibonaarm      & GNU 8.2.0      & OpenMPI 3.1.2  \\
\dibonaarm      & \arm 19.1      & OpenMPI 3.1.2  \\
\mn             & GNU 8.1.0      & Intel MPI 2018.4  \\
\mn             & Intel 2019.4   & Intel MPI 2018.4
\end{tabular}
\label{tabGraphSoftwareEnvironments}
\end{table}

\subsubsection{Node-to-node comparison}

To compare the computing and communication capabilities of the \txtwo and \skylake processors under an irregular workflow, we executed the MPI-only version of the \graph benchmark on a single node of \dibonaarm and \mn.

\begin{figure}[htbp]
  \includegraphics[width=\columnwidth]{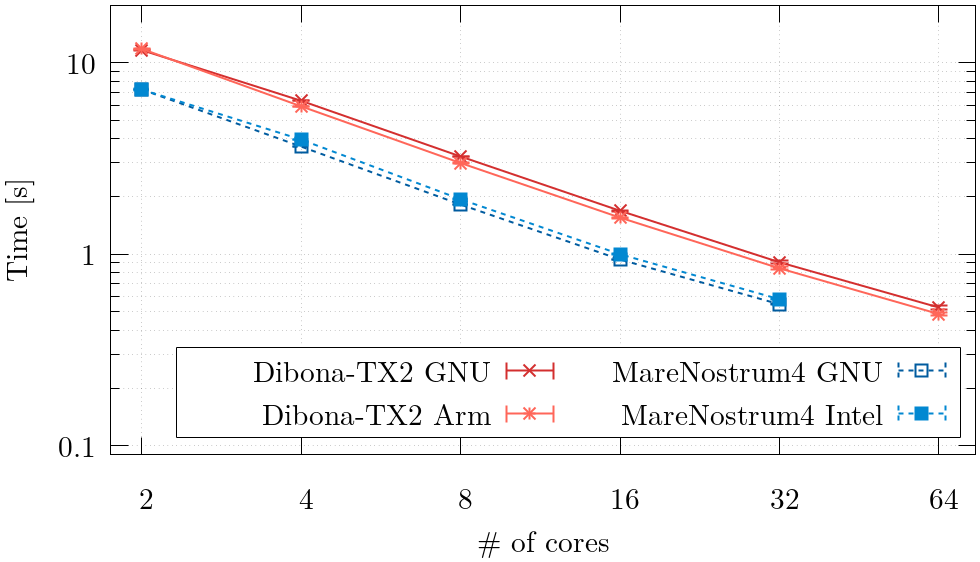}
  \caption{\graph -- Strong scaling on \dibonaarm and \mn with power of 2 ranks, with execution time as a function of number of cores}
  \label{figGraphDibonaStrongScalingTimePow2}
\end{figure}

Figure~\ref{figGraphDibonaStrongScalingTimePow2} shows the average time of a BFS algorithm varying the number of MPI processes, when compiling the benchmark optimized to run on a number of MPI processes that is a power of 2. Notice that, in this scenario, we are not able to use all the cores of one node of \mn. We can see that the performance obtained by the different compilers within the same architecture is very similar; the difference in all the cases is below $8\%$. Notice also that, in a core-to-core comparison, \dibonaarm is between $60\%$ and $80\%$ slower than \mn when using the GNU compiler suite and between $45\%$ and $60\%$ slower when comparing vendor-specific compilers. Nevertheless, when using the full node, \dibonaarm is $\sim4\%$ faster than \mn when using the GNU compiler suite and $\sim16\%$ when using the vendor-specific compilers.

\begin{figure}[htbp]
  \includegraphics[width=\columnwidth]{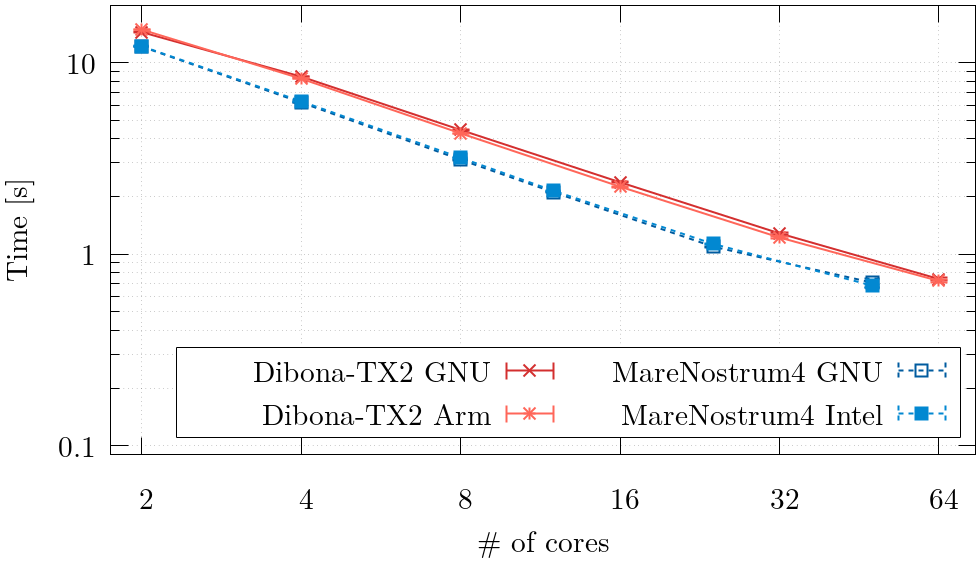}
  \caption{\graph -- Strong scaling on \dibonaarm and \mn with non-power of 2 ranks, with execution time as a function of number of cores}
  \label{figGraphDibonaStrongScalingTimeNotPow2}
\end{figure}

In Figure~\ref{figGraphDibonaStrongScalingTimeNotPow2}, we depict the same metric when the MPI processes are not a power of two. The performance delivered by different compilers in \dibonaarm is below 5.5\% in all the cases and for \mn below 3.5\%. When compiling with the GNU compiler suite and in a core-to-core comparison, \dibonaarm is between $12\%$ and $40\%$ slower than \mn, and between $5\%$ and $35\%$ when using vendor-specific compilers. When using the full node, \dibonaarm is $\sim4.5\%$ slower with GNU and $\sim6\%$ with vendor-specific compilers.

As expected, the benefit of using a binary optimized to use a power of two number of MPI processes implies a non-negligible performance gain. We measured a benefit between 25\% and 40\% on \dibonaarm with both compilers and between $60\%$ and $70\%$ on \mn. If we compare the performance of using a full node of \mn, running the power of 2 version with 32 cores is $30\%$ faster with GNU and $18\%$ faster with the Intel compiler than running with 48 cores with the non-power of 2 version.

\begin{figure}[htbp]
  \includegraphics[width=\columnwidth]{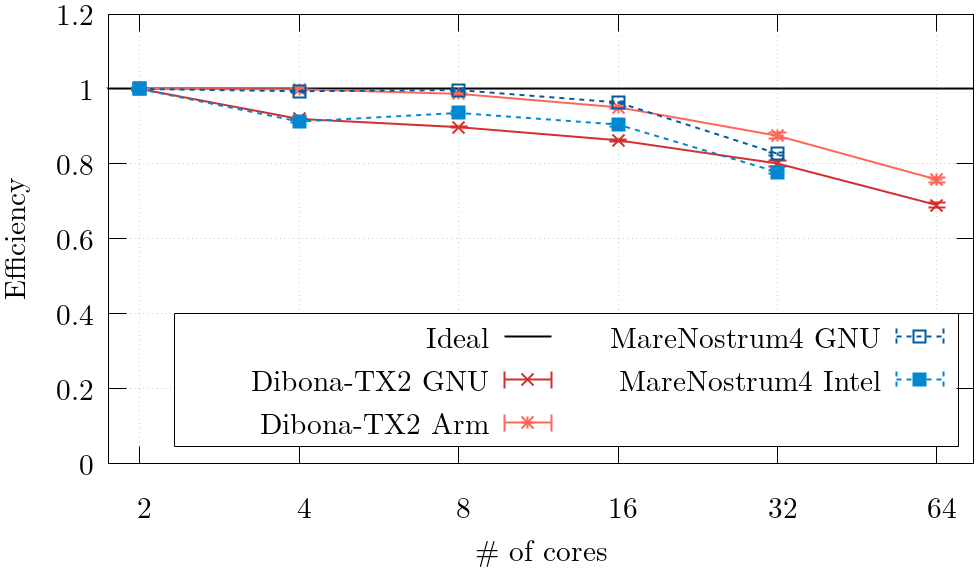}
  \caption{\graph -- Strong scaling on \dibonaarm and \mn with power of 2 ranks, with parallel efficiency as a function of number of cores}
  \label{figGraphDibonaStrongScalingEffPow2}
\end{figure}

In Figure~\ref{figGraphDibonaStrongScalingEffPow2}, we can see the parallel efficiency obtained in \mn and \dibonaarm when using different compilers and running with a power of 2 number of ranks. We can see that the parallel efficiency in both architectures have a similar trend, it drops when using 32 cores, probably due to memory bandwidth. This also explains that the drop in \mn is higher than in \dibonaarm because \dibonaarm offers a higher memory bandwidth.

\begin{figure}[htbp]
  \includegraphics[width=\columnwidth]{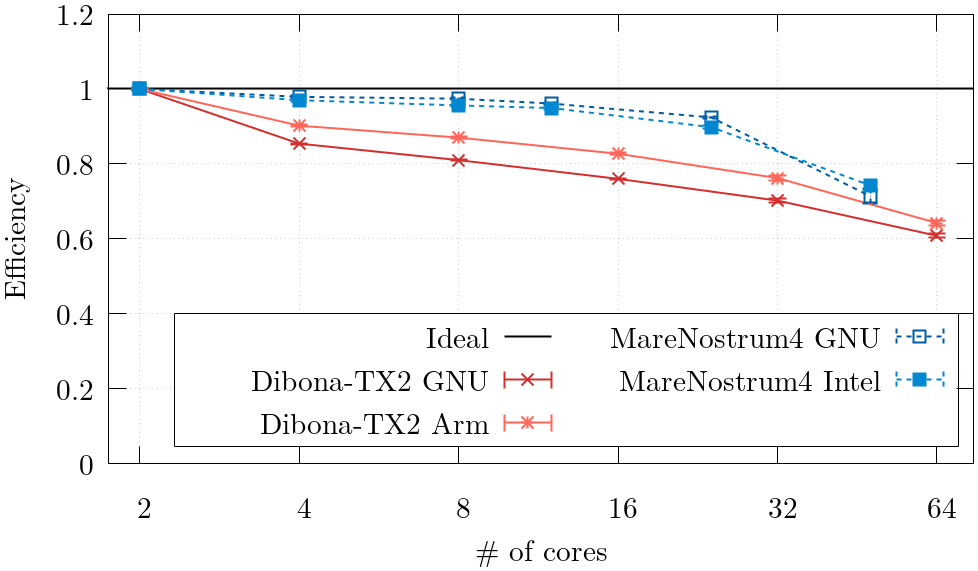}
  \caption{\graph -- Strong scaling on \dibonaarm and \mn with non-power of 2 ranks, with parallel efficiency as a function of number of cores}
  \label{figGraphDibonaStrongScalingEffNotPow2}
\end{figure}

In Figure~\ref{figGraphDibonaStrongScalingEffNotPow2}, we can find the same data plotted for a non-power of 2 number of MPI ranks. In this case, we observe that the use of a non-optimized code is affecting \dibonaarm more than \mn. Nevertheless, we observe the same effect of parallel efficiency dropping when using more than 32 cores and the fact that it has a higher impact on the parallel efficiency of \mn.

\section{Energy Considerations}\label{secEnergy}

In this section, we report energy measurements intending to compare the energy efficiency of state-of-the-art HPC architectures based on \arm and x86 CPUs. To this aim, we measured the energy consumption of our workloads on \dibona's \txtwo (\dibonaarm) and \skylake (\dibonaintel) nodes. 
We do not perform an energy/power comparison between \dibona and \mn because the power monitoring infrastructure of the two systems is significantly different, and the measurements would be neither homogeneous nor comparable. 
In Dibona, the power monitoring infrastructure allows us to homogeneously measure the power consumption of the whole motherboard of both types of compute nodes (\dibonaarm and \dibonaintel). However, in \mn, the power monitoring is based on the on-chip RAPL counters, which do not take into account the static power drain of the boards and the on-board components.

Our method has the following unique strengths: 
{\em i)} 
we employ complex production codes instead of benchmarks;
{\em ii)} 
since the system integrator of the \dibona system (ATOS/Bull) is the same for nodes powered by \arm and x86, these share a common power monitoring infrastructure, and the location of energy consumption sensors within the system is the same so that we can assure fair measurements;
{\em iii)} 
as we employ a production system, we can collect energy figures as users, testing the actual accessibility to the energy information, and providing data gathered on a final production system rather than on a specialized test bench.

The system also offered the possibility of measuring the {\em Energy To Solution} (E2S) of complete runs, as well as portions of code with no impact on the performance of the application monitored. We consider, therefore, the energy-to-solution as the relevant metric for our study in this section. We also present the {\em Energy Delay Product} (EDP) (\ie the product of the time to perform a run and its energy consumption) as a metric to combine performance and power drain over time. Despite its convenience, EDP, like E2S, does not allow to compare directly different applications since it depends on the execution time. The reader interested in a metric related to the energy efficiency (independent on the execution time) should take a closer look at LBC, which offers the MLUP/s and the MLUP/J metrics defined in Section~\ref{secLBC}. A summary of the energy measurements for all applications is given in Table~\ref{tab:energy}.
Note that these energy measurements relate to the whole node, including the CPU, the memory, the network interfaces, the I/O devices, and most of the motherboard. It is, therefore, not possible to make strong statements about the relative energy efficiency of the underlying processor architecture. 
Unless otherwise noted, we use the same experimental setup, as described in the previous section. 

\begin{table}[htbp]
\centering
\caption{Summary of energy measurements in one node}
\resizebox{\columnwidth}{!}{
\rowcolors{2}{gray!15}{white}
\begin{tabular}{l|c|c|c|c}
                                        & \multicolumn{2}{c|}{\bf \dibonaarm}               & \multicolumn{2}{c}{\bf \dibonaintel}                          \\
\midrule
  Compiler                              & GNU                     & Arm                     & GNU                        & Intel           \\
  {\bf Alya}                            &                         &                         &                            & \\
  Simulation time [s]                   & 347.40                  & 223.81                  & 236.13                     & 151.76        \\
  E2S    [kJ]                           & 90.17                   & 63.44                   & 101.12                     & 69.24         \\
  EDP    [kJs]                          & 31325.1                 & 14198.5                 & 23877.5                    & 10507.9       \\
  {\bf LBC}                             &                         &                         &                            & \\
  Simulation time [s]                   & 251.64                  & 333.99                  & 205.87                     & 208.53 \\
  Performance \si{[MLUP/\second]}       & 266.7                   & 200.9                   & 285.2                      & 281.6 \\
  E2S    [kJ]                           & 82.20                   & 107.09                  & 95.61                      & 97.89 \\
  Energy eff. [\si{MLUP/\joule}]        & 0.82                    & 0.63                    & 0.61                       & 0.60 \\
  EDP    [kJs]                          & 20681.5                 & 35767.4                 & 19683.5                    & 20413.2 \\
  {\bf Tangaroa}                        &                         &                         &                            & \\
  Simulation time [s]                   & 101.16                  & 93.77                   & 55.84                      & 61.20         \\
  E2S    [kJ]                           & 27.38                   & 25.17                   & 24.78                      & 27.69         \\
  EDP    [kJs]                          & 2769.76                 & 2360.19                 & 1383.72                    & 1694.62       \\
  {\bf \graph (pow of 2)}             &                         &                         &                            & \\
  Simulation time [s]                   & 39.84                 & 37.34                   & 53.80                       & 48.53\\
  E2S    [kJ]                           &12.00                  & 11.29                   & 17.69                       & 16.00\\
  EDP    [kJs]                          &477.98                 & 421.37                  & 951.48                      & 776.30\\
  {\bf \graph (generic)}              &                         &                           &                           & \\
  Simulation time [s]                   & 54.79                 & 52.61                     & 49.33                     & 46.46\\
  E2S    [kJ]                           & 15.66                 & 14.95                     & 21.20                     & 19.95\\
  EDP    [kJs]                          & 857.80                & 786.60                    & 1045.85                   & 926.76\\
\bottomrule
\end{tabular}
}
\label{tab:energy}
\end{table}

\subsection{Alya}

For Alya, we report in Table~\ref{tabAlyaHeatmapEnergy} the E2S in $kJ$ of 10 time steps of the respiratory simulation introduced in Section~\ref{secAlya}. 
Each column indicates the energy consumption when running a binary generated either with the GNU compiler suite or with vendor-specific alternatives, the  \arm HPC compiler for \dibonaarm, and the Intel suite for x86 \dibonaintel.
The color code indicates in green the lowest E2S and in red the highest E2S.
\begin{table}[htbp]
  \centering
  \includegraphics[width=\linewidth]{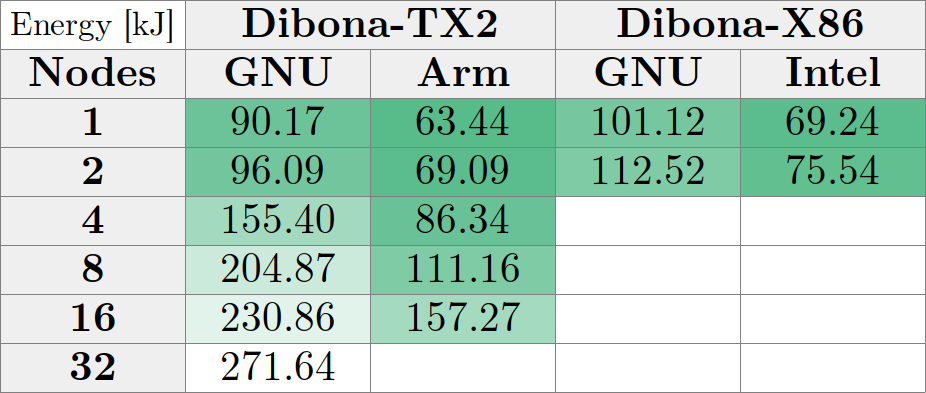}
  \caption{Alya: Energy to solution (in $kJ$) comparison among different compilers/architectures}
  \label{tabAlyaHeatmapEnergy}
\end{table}

As highlighted when analyzing the performance in Section~\ref{secAlyaNodeToNodeComp}, the vendor-specific compilers allow one to solve the same problem with less energy.
It is interesting to see that the compute nodes of \dibonaarm obtain the lowest E2S, although they are not faster than the \dibonaintel ones. This observation is valid for both generic and vendor-specific compilers, \txtwo-based nodes are more efficient than \skylake-based ones when comparing against the same kind of compiler. 
In Table~\ref{tabAlyaHeatmapTime}, the reader can compare the elapsed time in~$s$ for the same Alya case with the same color code used in Table~\ref{tabAlyaHeatmapEnergy} applied to the execution time (green = faster; red = slower).
Analyzing the performance at scale, it is interesting to note that running the same scientific simulation with the \arm HPC Compiler is 10\% faster than using GFortran. Still, its overall energy consumption is 30\% lower with the vendor-specific compiler than with the GNU compiler suite.
\begin{table}[htbp]
  \centering
  \includegraphics[width=\linewidth]{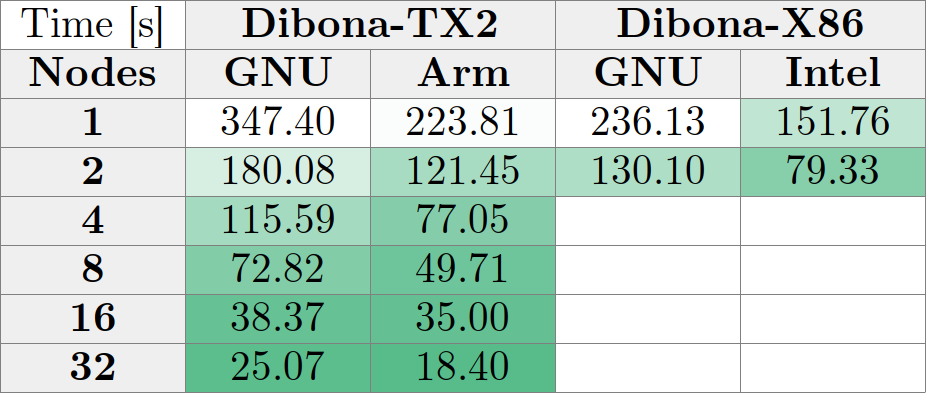}
  \caption{Alya: Execution time (in $s$) comparison among different compilers/architectures}
  \label{tabAlyaHeatmapTime}
\end{table}

As described in Section~\ref{secAlya}, Alya is a complex scientific code composed of several phases with different computational characteristics (see Table~\ref{tabAlyaArithIntensity} for details).
Focusing on one node, Table~\ref{tab:energy} shows that, even if \skylake processors are $\sim\!30\%$ faster than \txtwo, the energy consumption is slightly smaller on the \dibonaarm nodes. Because of this, the \dibonaintel nodes end up being $\sim\!30\%$ more efficient. It is very likely, however, that from the energy point of view, the simulation we studied can take more advantage of the 30\% higher memory bandwidth, offered by the architectural choice implemented by Marvell in the \txtwo, than the $4\times$ wider SIMD unit provided by the x86 \skylake CPU. 

Moreover, it is worth mentioning that this advantage of \skylake is lost if we compare the execution with the GNU compiler on \dibonaintel nodes with the \dibonaarm nodes using the \arm compiler. The time-to-solution is very similar, but the energy consumption and the efficiency are $\sim\!40\%$ better in the \dibonaarm nodes.

\subsection{LBC}
\label{secLBCenergy}

LBC has a relatively long
initialization phase, which is disregarded in the application's performance
measurement. For simplicity, we chose to read out the energy consumption counters
only at the beginning and the end of the application.
To decrease the impact of the initialization phase on the energy
reading, we increased the number of time steps from 10 to 500 for these
measurements. We expected that the initialization phase accounts for less than
a few percent of the energy consumption, which has been confirmed by varying
the number of time steps.  Note that the error reported below is the statistical
standard deviation of the measurement sample and does not include this
initialization bias.

In Table \ref{tab:energy}, we have chosen to report energy efficiency in terms
of lattice updates (\ie work done) per consumed energy unit. For the GNU compiler,
the energy efficiency is \SIUS{0.82+-0.01}{MLUP/\joule} on the \dibonaarm node,
while for \dibonaintel it is \SIUS{0.61+-0.01}{MLUP/\joule}.
The results show that, for
this particular code, the energy efficiency (in terms of the domain-specific
metric \si{MLUP/\joule}) of \dibonaarm is roughly \SI{30}{\percent} higher
than that of \dibonaintel.

In addition to the GNU compiler, we   also used the compilers of the
respective processor vendor, \ie Arm and Intel. On \txtwo, the Arm compiler
produces significantly lower performing and energy-efficient code than the GNU
compiler. On \skylake, the Intel compiler, on the other hand, yields practically
identical performance and energy efficiency compared with the GNU compiler.

\subsection{Tangaroa}

As with the execution times, the energy measurements of Tangaroa are constrained to the simulation phase. These are summarized in Table~\ref{tab:energy}, and it can be seen that for each platform, the energy consumption is almost proportional to the execution time. This is a consequence of the effort Tangaroa makes to hide communication delays and achieve a sustained IPC. Therefore, the compilers that give the best time also improve the energy consumption. Between the two platforms, the table shows that, although the simulation time in a single node is 67\% longer for \dibonaarm, the energy consumed in the region of interest is only 1.5\% higher. This reflects on the efficiency values, making \dibonaintel nodes $\sim\!40\%$ more efficient than \dibonaarm ones.

\subsection{\graph}

For the \graph benchmark, we report figures of energy (E2S and EDP) in Table~\ref{tab:energy} when performing 64 BFS operations with scale 24 for both cases, when we run with power of 2 MPI processes and when we run with a generic number of MPI processes.
As highlighted for Alya, also in the \graph benchmark, vendor-specific compilers deliver more performance, wasting less energy for a given problem.

The reader should also note the higher performance and higher energy efficiency of \dibonaarm when running with power of 2 MPI processes.
The fact that \dibonaarm nodes house a number of cores that are a power of 2 allows us to run using all cores of the node, while in \dibonaintel, we run 32 processes on each node, leaving 24 cores idle.
Still, even taking this into account, \dibonaarm nodes are also 30\% more energy-efficient when running a non-power of 2 number of MPI processes.

\section{Scalability}\label{secScalability}

The evaluation of the \dibonaarm platform has been performed so far on a small scale. However, \arm-based compute nodes are being considered as building blocks of large systems to progress towards Exascale computing. Therefore, we analyze in this section our set of applications in a multi-node context, as we think that a study at the scale of a thousand cores reveals valuable insights for extrapolating performance for larger systems.

In this section, the speedup sample points coincide with full nodes instead of cores. The efficiency is then computed as $E = \frac{t_1}{t_i \cdot i}$, where $t_1$ is the execution time when running with one node, and $t_i$ is the execution time when running with $i$ nodes.

The reader should note that we perform our scalability tests scaling up the number of compute nodes disregarding the number of cores housed in each node.
The reason for that is twofold: 
on the one hand, we have already presented in Sections~\ref{secHw} and~\ref{secEval} the performance of a single core/node of the \dibonaarm cluster; 
on the other hand, we aim at providing insightful information to domain scientists and HPC facility managers interested in the \dibonaarm technology, being well aware that the basic unit when acquiring/deploying a cluster is a single compute node.

Since \dibonaintel has only three \skylake nodes, we have resorted to \mn to execute the applications with many nodes. However, the software stack used in the \skylake experiments is still Intel 2018. Unless otherwise noted, we use the same experimental setup, as described in the previous sections.

\subsection{Alya}

\begin{figure}[htbp]
  \centering
  \includegraphics[width=\linewidth]{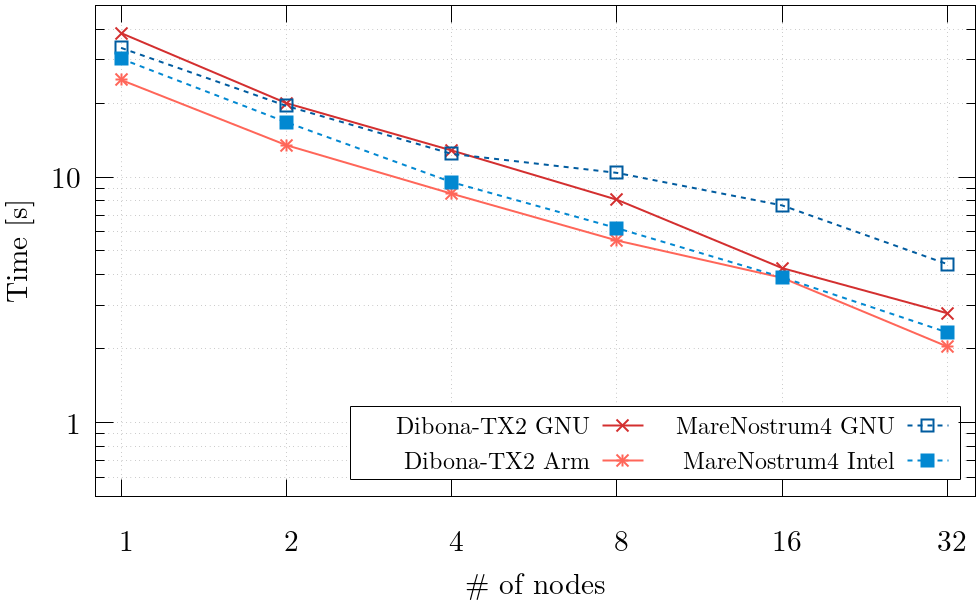}
  \caption{Alya -- Strong scalability of the execution time on \dibonaarm and \mn using different compilers}
  \label{figScalability}
\end{figure}

Figure~\ref{figScalability} shows the execution time per time step in seconds for both node types, also with the generic and vendor software stacks mentioned above.
We can observe that, in all the cases, the vendor-specific compiler outperforms the generic one. The solution that delivers the best performance is the \dibonaarm cluster with the \arm HPC compiler, closely followed by less than 10\%, is the Intel version in \mn.
We can also see that the two versions using the generic GNU compiler perform worse. Up to four nodes, their performance is very similar, but for more than four nodes, the performance of the GNU version in \mn drops. We do not have a clear explanation for this, but we suspect that the different cache microarchitecture of the \txtwo and the \skylake CPU combined with the workload distribution performed by Alya can produce this kind of behavior using the GNU compiler toolchain at a mid-large number of MPI processes.

\begin{figure}[htbp]
  \centering
  \includegraphics[width=\linewidth]{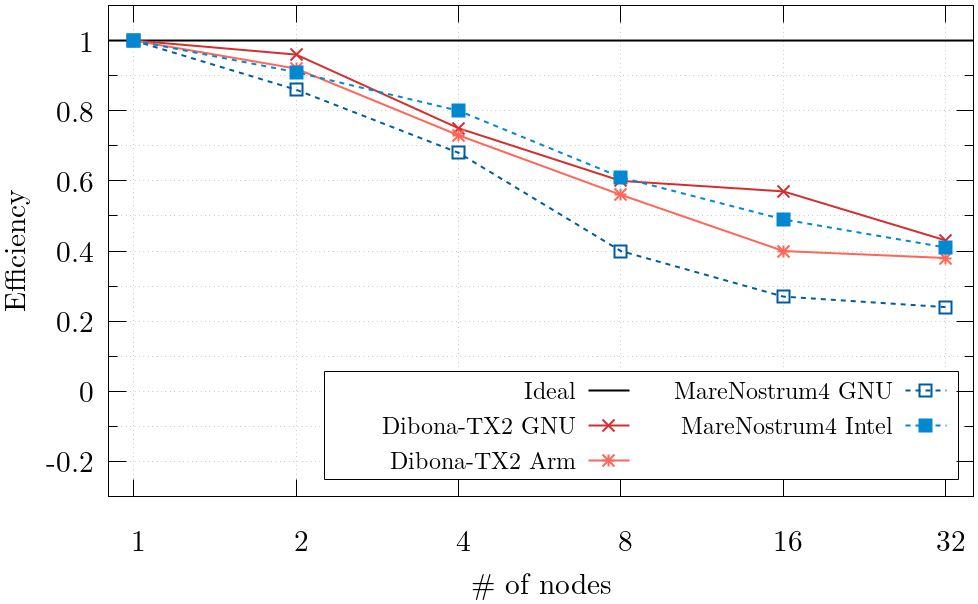}
  \caption{Alya -- Parallel efficiency of Alya \dibonaarm and \mn using different compilers}
  \label{figEfficiency}
\end{figure}

In Figure~\ref{figEfficiency}, we show the parallel efficiency obtained by each run.
The GNU compiler on \mn appears to be the less efficient version of our test cases, reaching an efficiency of 24\% when running with 32 nodes of \mn. The performance drift when passing from four to eight nodes seems to be the root cause of this bad scalability.
On \dibonaarm, the run with the Arm compiler is the one obtaining worse parallel efficiency, although Figure~\ref{figScalability} shows that this is also the fastest case.
This is a common effect when computing metrics over different references.

Moreover, we can see that the parallel efficiency of the \arm compiler version in \dibonaarm is slightly worse than the one obtained by the Intel compiler in \mn. This can be an effect of the configuration of the software layers leveraging the underlying physical layer of the network. We have seen, in fact, in Section~\ref{secNetwork}, that the \dibonaarm network still suffers some configuration glitch.

Comparing the runs with GNU compilers in both platforms, we observe again that the \dibonaarm cluster obtains a better performance, also resulting in a better parallel efficiency. 

\subsection{LBC}\label{secLBCscal}

The code LBC is intended to be used for large problems, and at the time of data
acquisition, only 16 \txtwo nodes were regularly available in \dibonaarm. In
order to study scalability across nodes, we have therefore decided to perform a weak
scaling experiment. Strong scaling would be expected to become necessary at
larger scales only. We compare results obtained on \dibonaarm and \mn.

\begin{figure}[htbp]
  \centering
  \includegraphics[width=\linewidth]{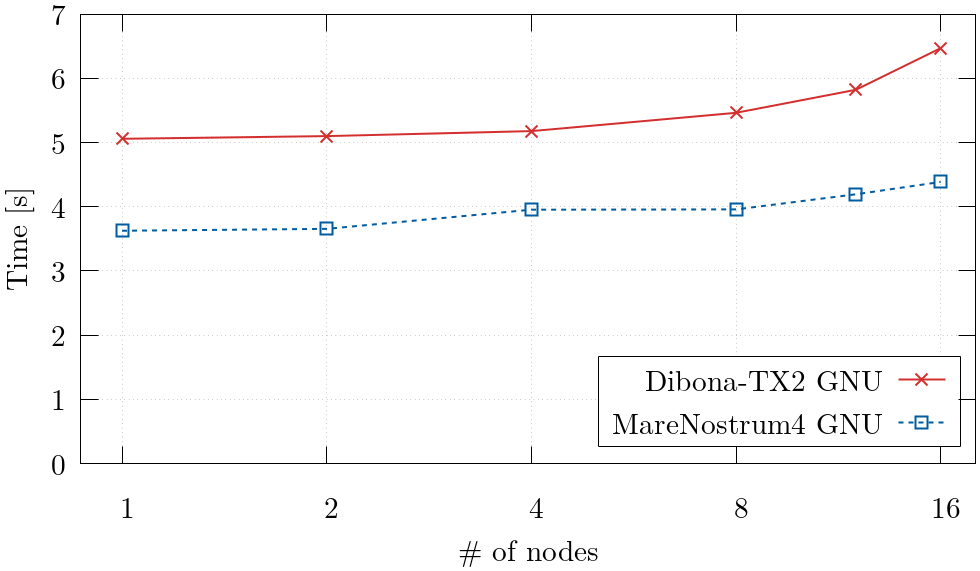}
  \caption{ LBC -- Weak scaling on \dibonaarm and \mn }
  \label{figScalabilityLBC_time}
\end{figure}

For the weak scaling experiment, we have kept the problem size per node $n_N$
equal to the single-node runs presented above (see Table~\ref{tab:LBC_1node}).
We performed at least 30 executions of LBC on each machine. To get a
representative result, at most 10 runs were done within the same job, and jobs
were distributed over two days, in most cases. Again, we use the application-specific
metric \si{MLUPS} as a proxy for performance. 
Unless stated otherwise, the statistical variation of measurements is at most
\SI{1}{\percent} on \mn. For \dibonaarm, it is between \SI{1}{\percent}
and \SI{5}{\percent}.

\begin{figure}[htbp]
  \centering
  \includegraphics[width=\linewidth]{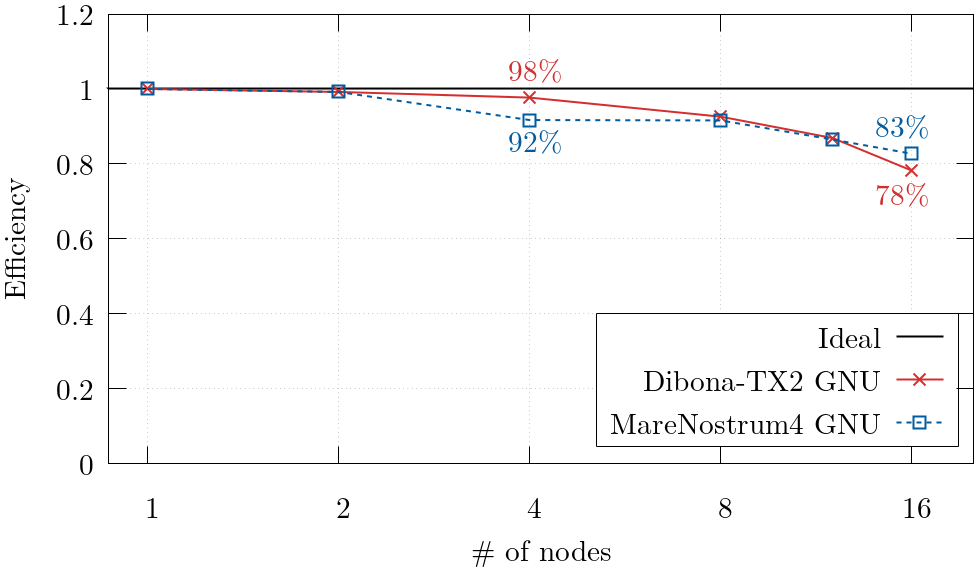}
  \caption{ LBC -- Efficiency with respect to the baseline of one node as a function of nodes on \dibonaarm and \mn }
  \label{figScalabilityLBC_eff}
\end{figure}

The results of a weak scaling experiment on \dibonaarm and \mn are
illustrated in Figure~\ref{figScalabilityLBC_time}, where we plot the weak scaling
time, and Figure~\ref{figScalabilityLBC_eff}, where we represent the efficiency with respect to a single node.

When increasing the number of nodes from 1 to 16, the scaling efficiencies on
\dibona and \mn drop steadily. However, the slope is slightly larger for
\dibona than for \mn. At 16 nodes, \dibona ends up at around \SI{78}{\percent} scaling
efficiency, while \mn achieves \SI{83}{\percent}. However, this difference of
\SI{5}{\percent} is only marginally significant when taking into account
the measurement error. 
 
The dip at 4 nodes for \mn is due to a series of closely spaced runs that show
substantially higher execution times. Closer inspection of the logs revealed an
increased communication time, presumably due to a particularly high intermittent
network load. Disregarding these runs moves the data point right on top of a
gradually sloping line.   
  
\subsection{Tangaroa}

\begin{figure}[htbp]
   \centering
   \includegraphics[width=\columnwidth]{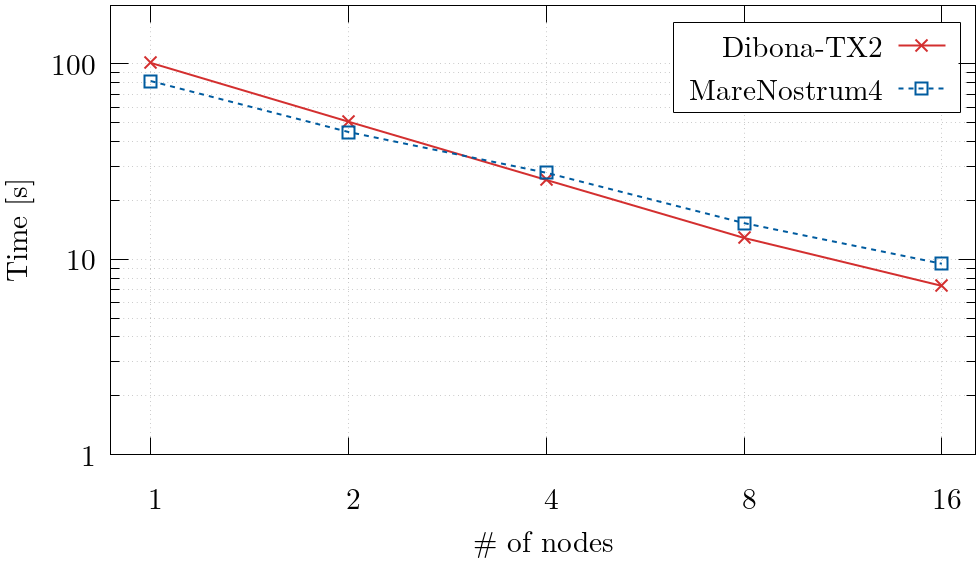}
   \caption{Tangaroa -- Strong scalability of the execution time on \dibonaarm and \mn}
   \label{tangaroa_scaling_time}
\end{figure}

\begin{figure}[htbp]
   \centering
   \includegraphics[width=\columnwidth]{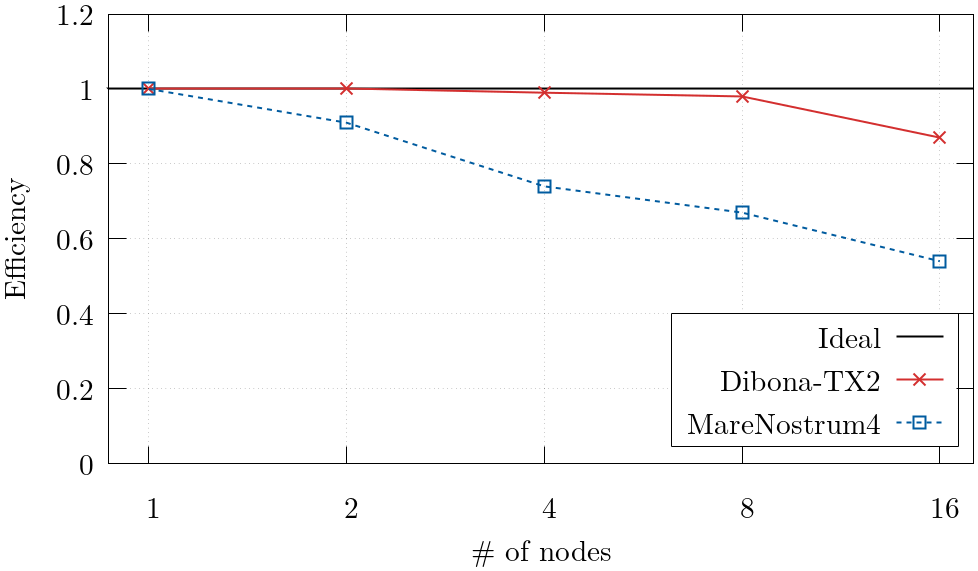}
   \caption{Tangaroa -- Parallel efficiency on \dibonaarm and \mn}
   \label{tangaroa_scaling_eff}
\end{figure}

A set of strong-scaling experiments of Tangaroa were made using GCC as the compiler in both platforms, \dibonaarm and \mn. The execution times of these experiments are shown in Figure~\ref{tangaroa_scaling_time}, while the parallel efficiency is depicted in Figure~\ref{tangaroa_scaling_eff}. The base times of the efficiency plot for the single node executions appear in Table~\ref{tangaroa_node}. 

It is apparent that the scalability is far better in \dibonaarm. Note that the efficiency of the 16 node execution in \mn is close to 50\%, despite the nodes and the network being superior to \dibonaarm. The reason behind this effect is twofold. First, the size of the problem is not big enough for such a large number of processes, leading to computation bursts in the range of 100~ms. Second, since \mn is a production machine, network traffic is significantly higher than in \dibonaarm. As a consequence, the transmission delays are in the same order of magnitude as the computation bursts, and Tangaroa is less capable of hiding them.

\subsection{\graph}

For the scalability study with \graph, we use a size $s=27$, corresponding to a number of vertices $v = 2^{27}$. All the numbers reported are the duration of one BFS operation averaging 64 measurements, as reported by the application.

\begin{figure}[htbp]
   \centering
   \includegraphics[width=\columnwidth]{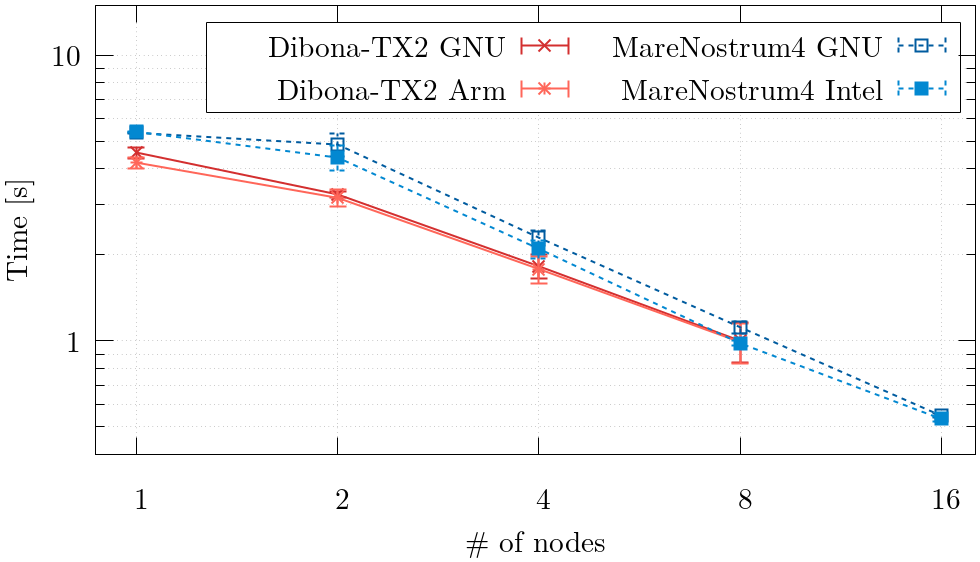}
   \caption{\graph -- Strong scalability of the execution time on \dibonaarm and \mn, power of 2}
   \label{graph_scaling_time_pow2}
\end{figure}

In Figure~\ref{graph_scaling_time_pow2}, we can see the average elapsed time to perform one BFS when running with a number of MPI ranks that is a power of 2. We can observe that up to 8 nodes \dibonaarm is faster than \mn. In both cases, the vendor-specific compiler is the one that delivers the best performance, but the performance gain with respect to GNU is below 10\% in all the cases. Although \dibona outperforms \mn in all the cases, the distance between them decreases as the number of nodes increase, being equal for 8 nodes.

\begin{figure}[htbp]
   \centering
   \includegraphics[width=\columnwidth]{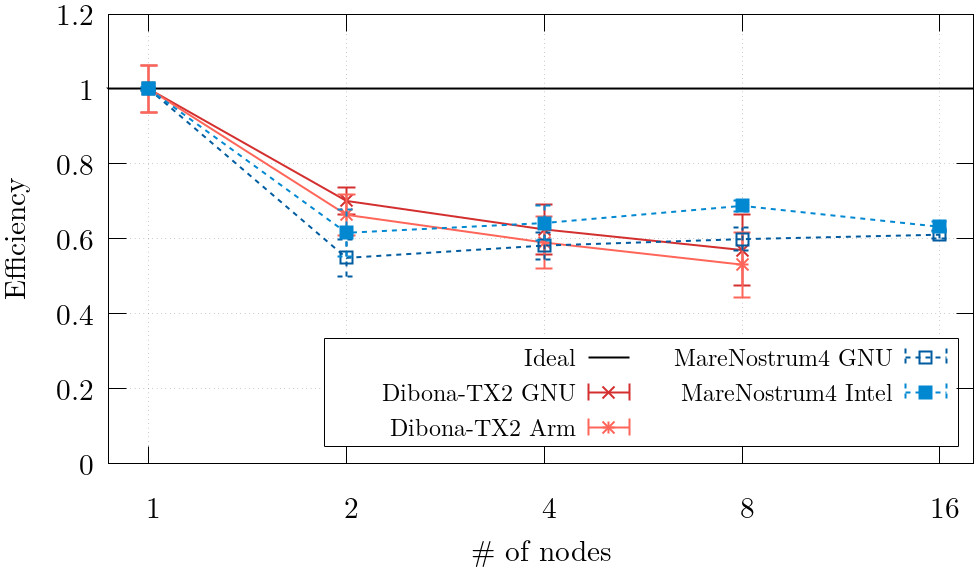}
   \caption{\graph -- Parallel efficiency on \dibonaarm and \mn, power of 2}
   \label{graph_scaling_eff_pow2}
\end{figure}

Figure~\ref{graph_scaling_eff_pow2} shows the parallel efficiency when running with a binary optimized to run in a number of MPI ranks power of 2. We can see that all the parallel efficiencies drop from 1 to 2 nodes. This confirms that \graph is network intensive because the parallel efficiency drops when the communications are done outside the node.

\begin{figure}[htbp]
   \centering
   \includegraphics[width=\columnwidth]{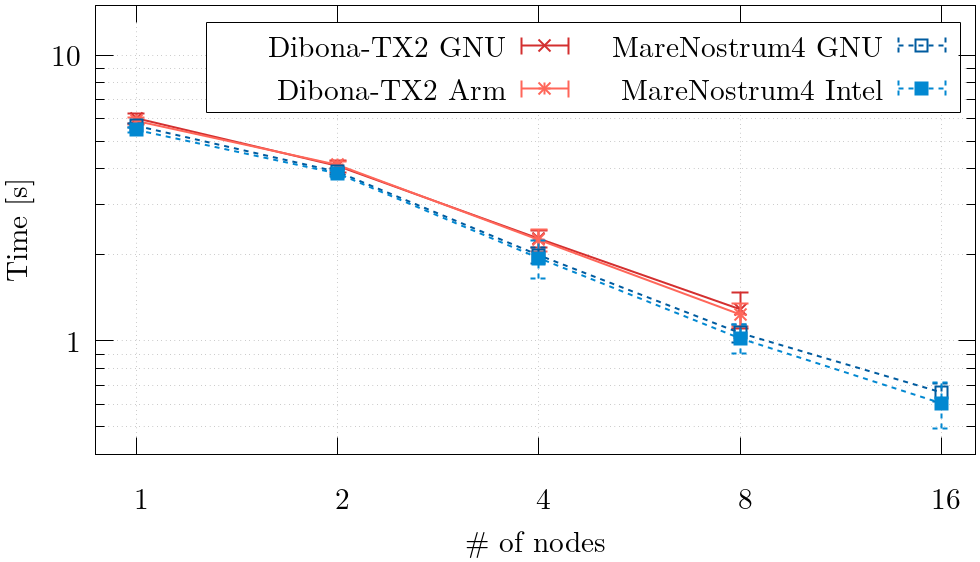}
   \caption{\graph -- Strong scalability of the execution time on \dibonaarm and \mn, not power of 2}
   \label{graph_scaling_time_not_pow2}
\end{figure}

In Figure~\ref{graph_scaling_time_not_pow2}, we show the execution time of one BFS when using a binary compiled to run in a non-power of 2 number of MPI processes. In this case, \mn outperforms \dibona. This difference in performance can be explained because, as we have seen in the previous sections, the compilers at \mn can obtain more performance from the non-optimized version of the code. It is important to notice that in this case, both clusters can use all the cores of each compute node.

\begin{figure}[htbp]
   \centering
   \includegraphics[width=\columnwidth]{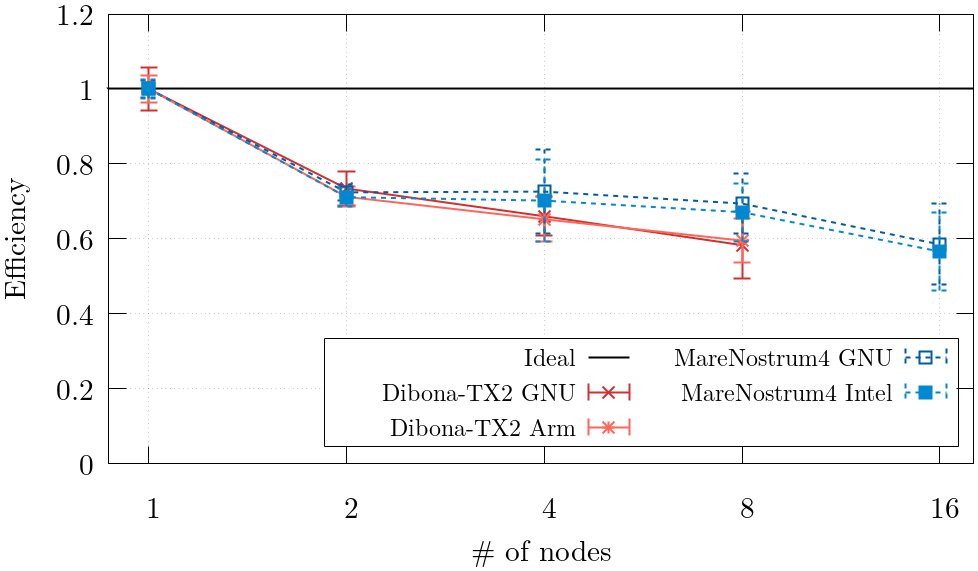}
   \caption{\graph -- Parallel efficiency on \dibonaarm and \mn, not power of 2}
   \label{graph_scaling_eff_not_pow2}
\end{figure}

Figure~\ref{graph_scaling_eff_not_pow2} depicts the parallel efficiency of the same executions. We can see that the big drop in performance when communicating outside a compute node is still present, meaning that, in this version, the communications are also an important factor.

\section{Scalability Projections}\label{secProjections}

\subsection{\graph Communication Study}

For analyzing further the behavior at scale of the \graph benchmark, we measured $t_{MPI}$, the time spent on the MPI calls, and $t_{Cal}$, the time spent   performing the local BFS operations.
The goal is to study how $t_{MPI}$ evolves, adding more processes, and to extrapolate its impact at scale.

Due to the irregular nature of the benchmark, we refined our model differentiating $t_{MPI}$ into $t_{Com}$, which is the time during which actual communications happen, and $t_{LB}$, which is the time due to the load balance across different MPI processes.
The total execution time $t_{TOT}$ can be therefore expressed as 
\begin{equation}\label{eqTmpiGraph}
t_{TOT} = t_{Cal} + t_{MPI} = t_{Cal} + t_{Com} + t_{LB}.
\end{equation}

The additive nature of Equation~\ref{eqTmpiGraph} allow us to study the ratio of each contribution with the total time:

\begin{equation}\label{eqratioTmpiGraph}
\frac{t_{Cal}}{t_{TOT}} + \frac{t_{Com}}{t_{TOT}} + \frac{t_{LB}}{t_{TOT}} = 1.
\end{equation}

\begin{figure}[htbp]
   \centering
   \includegraphics[width=\columnwidth]{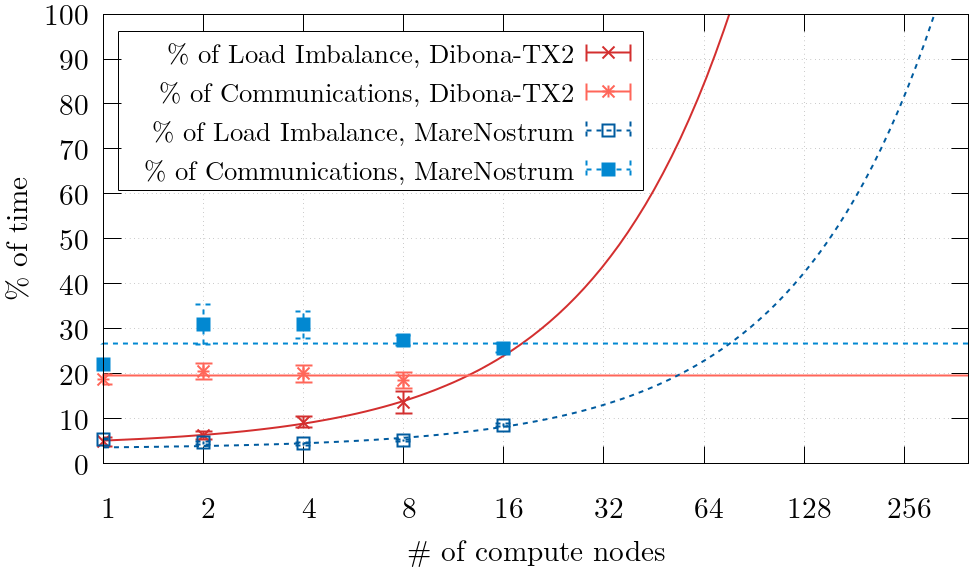}
   \caption{\graph -- MPI time on \dibonaarm and \mn, power of 2, vendor compilers}
   \label{graph_mpi_time}
\end{figure}

In Figure~\ref{graph_mpi_time}, we represent with points the percentage of time measured on the \dibonaarm and \mn when running \graph with scale 27 in the configuration with power of 2 MPI processes.
Since we are interested in   scalability, we studied the percentage of time spent in the communication (corresponding to $t_{Com}$) and the percentage of time spent due to load balance (corresponding to $t_{LB}$).
The solid lines are linear fits of those two contributions to $t_{MPI}$.
We interpolate the data of load balance with a linear function $ t_{LB} / t_{TOT} = a\cdot p + b$, where $p$ is the number of MPI processes, and $a$ and $b$ are the two parameters to fit. The rate of time spent in communication is modeled as a constant function $t_{Com} / t_{TOT} = c$. In Table~\ref{tabGraphParamFitMpi}, we present the parameters resulting from the fit.

\begin{table}[htbp]
\centering
\caption{Summary of $a$, $b$, and $c$ parameters for modeling $t_{MPI}$ in \graph}
\rowcolors{2}{gray!15}{white}
\begin{tabular}{c|c}
 {\bf \dibonaarm}	& {\bf \mn} \\
\midrule
a =	$1.26 \pm 0.13$ & a= $0.31 \pm 0.06$ \\
b =	$3.86 \pm 0.48$	& b= $3.29 \pm 0.59$ \\
c =	$19.59 \pm 0.62$ &	c= $26.77 \pm 0.83$ \\
\bottomrule
\end{tabular}
\label{tabGraphParamFitMpi}
\end{table}

The first observation is related to the trends visible in Figure~\ref{graph_mpi_time}. The factor that limits the scalability is the load balance across MPI processes that, in our projections, should reach a critical point already with $\sim 70$ compute nodes on \dibonaarm and with $\sim 270$ nodes in \mn.

The second remark is related to the percentage of time spent in performing the actual communication. This contribution seems to be 30\% higher on \mn than on \dibonaarm.
This is caused by the different technologies used to interconnect the nodes on the clusters and the network congestion of \mn (which is an operational production cluster).

\subsection{Speedup Projections}

In this section, we use the scalability data presented in Section~\ref{secScalability} for each application to project the behavior at scale of the two clusters under study.
The goal is to clarify the behavior at scale of Dibona under HPC workloads.

For projecting the scalability, we study the speedup with respect to the performance of a single node of each cluster. 
In the case of strong scalability, we use   Amdahl's law. The speedup $s$ as a function of the number of processes $p$ is computed as
\begin{equation}
s(p) = \frac{1}{(1-a) + \frac{a}{p}} + b
\label{eqAmdahl}
\end{equation}

In the case of weak scalability, we use   Gustafson's law. The speedup $s$ as a function of the number of processes $p$ is computed as

\begin{equation}
s(p) = (1-a) + a\cdot p
\label{eqGustafson}
\end{equation}

In both equations, $a$ is a parameter that expresses the time rate that is taking advantage of the parallelization.
For the strong scaling, we also take into account a parameter $b$, which represents an approximation of the parallelization overhead.

For Alya, Tangaroa, and \graph, we fit the strong scalability data presented in Section~\ref{secScalability} with Equation~\ref{eqAmdahl}, while for LBC we project the behavior at scale using Equation~\ref{eqGustafson}. This allows us to find the values of $a$.
In Table~\ref{tabProjection}, we report the values of $a$ for each application when running on \dibonaarm and \mn using the GNU Compiler Suite and the proprietary compilers.

\begin{table*}[htbp]
\centering
\caption{Summary of $a$ and $b$ parameters for projection of scalability}
\resizebox{.8\textwidth}{!}{
\rowcolors{2}{gray!15}{white}
\begin{tabular}{l|c|c|c|c|c}
                    &     & \multicolumn{2}{c|}{\bf \dibonaarm}           & \multicolumn{2}{c}{\bf \mn}            \\
\midrule
  Compiler          &     & GNU                     & Arm                 & GNU                     & Intel                 \\
  {\bf Alya}        & a = & $0.960 \pm 0.004$       & $0.947  \pm 0.010$  & $0.892 \pm 0.023$       & $0.954  \pm 0.006$    \\
                    & b = & $-0.685 \pm 0.395$      & $-0.890 \pm 0.772$  & $-0.634  \pm 0.685$     & $-0.674  \pm  0.505$  \\
{\bf LBC}          & a = & $0.817 \pm 0.027$       & N/A                 & $0.839 \pm 0.014$       & N/A                   \\
{\bf Tangaroa}      & a = & $0.989 \pm 0.002$       & N/A                 & $0.946 \pm 0.004$       & N/A                   \\
                    & b = & $0.162 \pm 0.149$       & N/A                 & $-0.324 \pm 0.173$      & N/A                   \\
{\bf \graph}        & a = & $0.918 \pm 0.013$       & $0.905 \pm 0.014$   & $0.968 \pm 0.006$       & $0.970 \pm 0.003$     \\
                    & b = & $-0.554 \pm 0.174$      & $-0.604 \pm 0.176$  & $-1.252 \pm 0.330$      & $-0.950 \pm 0.157$    \\
\bottomrule
\end{tabular}
}
\label{tabProjection}
\end{table*}

\begin{figure}[htbp]
  \centering
  \includegraphics[width=\linewidth]{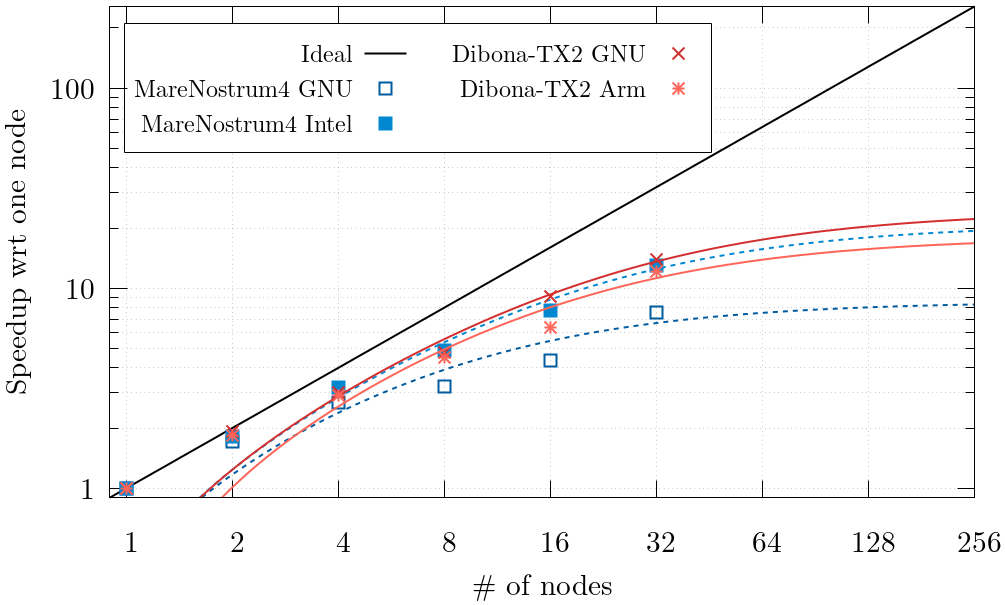}
  \caption{Alya -- Speedup projection of \dibonaarm and \mn using different compilers}
  \label{figAlyaProjection}
\end{figure}

\begin{figure}[htbp]
  \centering
  \includegraphics[width=\linewidth]{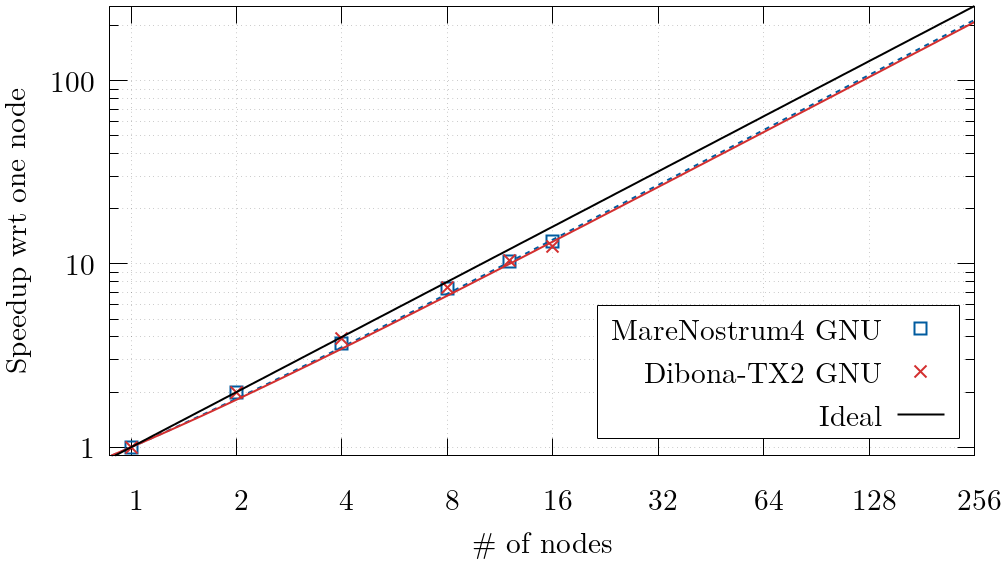}
  \caption{LBC -- Speedup projection of \dibonaarm and \mn using different compilers}
  \label{figLBCProjection}
\end{figure}

\begin{figure}[htbp]
  \centering
  \includegraphics[width=\linewidth]{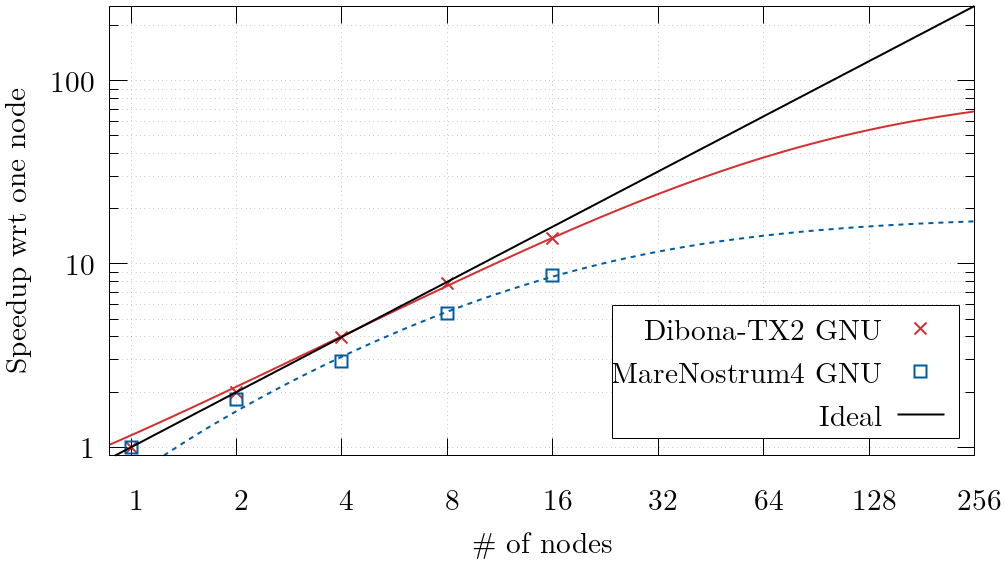}
  \caption{Tangaroa -- Speedup projection of \dibonaarm and \mn using different compilers}
  \label{figTangaroaProjection}
\end{figure}

\begin{figure}[htbp]
  \centering
  \includegraphics[width=\linewidth]{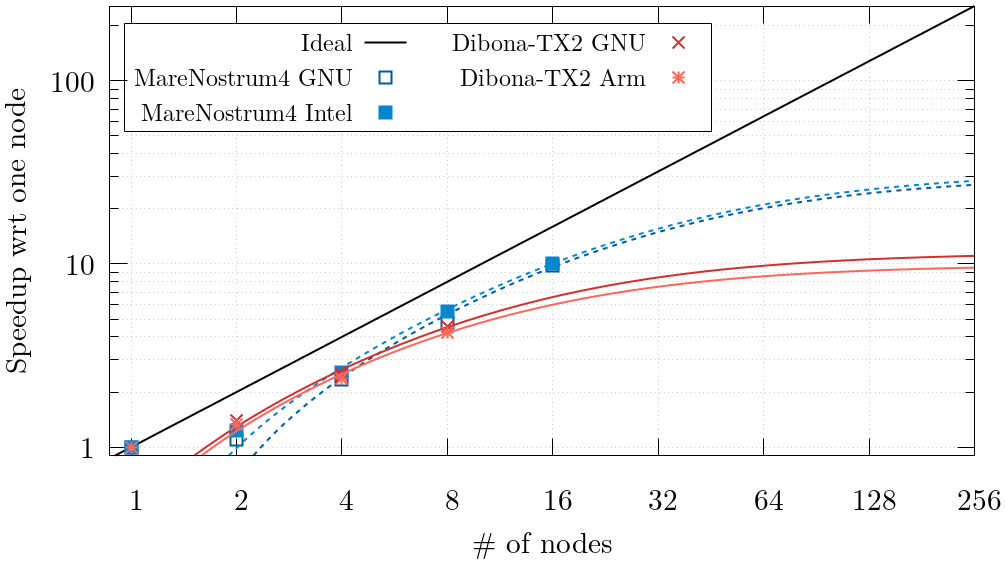}
  \caption{\graph (power of 2) -- Speedup projection of \dibonaarm and \mn using different compilers}
  \label{figGraphProjection}
\end{figure}

In Figures~\ref{figAlyaProjection},~\ref{figLBCProjection},~\ref{figTangaroaProjection}, and~\ref{figGraphProjection}, we present the projections at the scale of the four applications.
Points represent real measured data, while solid lines are projections obtained fitting the measurements with Equation~\ref{eqAmdahl} for Alya, Tangaroa, and \graph, and with Equation~\ref{eqGustafson} for LBC.

Looking at the strong scalability, we expect that the slope of the scalability curve becomes increasingly flat for all applications. This is an inherent problem of  splitting the workload across  an increasing number of computational units.
With our study, however, we demonstrate that \dibonaarm allows one to reach a scalability similar to \mn, even when projecting our measurements at a larger scale.
This means that the \dibonaarm technology can be considered as good as the one of \mn for large HPC clusters, besides the micro-architectural differences.

\section{Conclusions}\label{secConclusions}

In this paper, we analyzed the performance of the latest \arm-based CPU targeting the HPC market, Marvell (former Cavium) \txtwo. These processors were available to us in the \dibona cluster, developed within the European project, Mont-Blanc 3. For comparison, we performed the same experiments in state-of-the-art x86 \skylake processors in the same machine, as well as in the Tier-0 supercomputer \mn.

\subsection{CPU and system architecture}

We begin the analysis focusing on pure micro-benchmarking.
Perhaps the most salient fact about the \txtwo processor is the number of cores it integrates. With 32 cores per socket, a compute node of the \dibonaarm cluster has 33\% more than a \skylake-based node in \mn, our baseline machine, although their clock frequency is 5\% less than those in the \skylake running at 2.1~GHz.

Attending to the memory subsystem, the memory installed on the \dibonaarm nodes is DDR4-2666, which is 17\% slower than the DDR4-3200 coupled with the \skylake CPUs in \mn.
However, our experiments show that the \txtwo, with its eight memory channels per socket, gives a 25\% higher memory bandwidth than the \skylake, which only offers six channels.

The raw floating-point performance has been studied with a synthetic benchmark
in Section~\ref{secFloat}. Figure~\ref{fig:fpu} shows that both processors have scalar floating-point units
(FPUs) of similar characteristics. We have shown that the slight difference of
\SI{5}{\percent} in scalar floating-point performance can be explained by
different clock frequencies. 
The performance of the Intel AVX512 SIMD unit is four times higher than the
performance of the NEON SIMD unit found in the \txtwo CPU. Again, this is expected
as the AVX512 unit is four times wider than the 128-bit NEON SIMD unit.  

Summarizing, the \txtwo processors have more cores and memory channels than the \skylake, but slower clock and memory, and substantially shorter SIMD units.

\subsection{Software stack maturity}
An important aspect when considering new architectures, is the maturity of the software stack.
This includes not only the compiler but also libraries enabling parallel programming, such as MPI.
Our experiments show that, in general, the software stack provided by \arm has a high maturity level, comparable to that of the Intel tools and libraries. 
On the other side, the \arm support offered by the open-source stack is on par with Intel, although both usually show less ($\sim 30\%$) performance than the vendor-provided stacks.
This confirms the trend highlighted by F.~Banchelli et al. in~\cite{IsARMSoftware}.

\subsection{HPC applications performance and scalability}

While in the first part of the paper we focus on pure micro-benchmarking, in the second part we analyze the performance of three complex scientific applications at scale: Alya, LBC, and Tangaroa.
%
%
Alya is the most complex code of the set. It is a multi-physics solver based on finite-element methods.
%
%
LBC is an implementation of the Lattice Boltzmann method of solving fluid dynamics, showing a stencil-like data access pattern.
%
%
Tangaroa is a particle tracking code with single-precision arithmetic.

We also complement our study including the \graph benchmark, as a representative of emerging workloads that appear to be increasingly relevant in modern datacenters.

The four codes were executed within a single node of each machine.
The performance of the \dibonaarm nodes showed to be lower than the \skylake-based compute nodes, ranging from $\sim 4\%$ slower for LBC, to $\sim 30\%$ and $\sim 40\%$ slower for Alya and Tangaroa, respectively, which means that the extra number of cores and memory channels of the \txtwo CPU does not entirely overcome the limitations on the SIMD units.
However, further experiments within a single node showed that the applications scaled better in \dibonaarm than in \mn.
This can be attributed to the presence of two extra memory channels and a different cache micro-architecture.

Concerning scalability across nodes, the strong scaling experiment with Tangaroa on \dibonaarm gave scaling efficiencies of roughly \SI{85}{\percent}, which is substantially higher than on \mn, where the efficiency drops to approximately \SI{55}{\percent}.
Note that as expected from a strong scaling experiment, the problem size becomes progressively smaller, and 
the relative importance of MPI communication time increases. This explains the gradual decrease in scaling efficiency. Contrary to \dibonaarm, however, \mn is a production cluster with an extensive diameter network and a high number of users active at any time. Thus, network traffic by other users will, therefore, have a much higher effect MPI communication time on \mn than on \dibonaarm, which explains the worse scaling.    
However, for Alya, and more so for LBC,  the MPI communication is not as performance-critical as for Tangaroa, and the scaling efficiencies observed on both platforms are almost identical for LBC and are within 10\% of each other, respectively.
Our experiments with  synthetic MPI bandwidth benchmark (see Section~\ref{secNetwork}), however, show that at the time of our data acquisition, \dibonaarm's OpenMPI was misconfigured and reached a lower bandwidth for intermediate message sizes than expected from raw low-level network benchmarks. 
This shows that the Infiniband interconnection implemented in \dibonaarm is at the level of state-of-the-art HPC systems, even if the tuning of the OpenMPI stack leveraging the network is not optimal. 
Another important conclusion resulting from the study of the \graph benchmark is that for irregular workloads, the most important limiting factor at scale is the load balance across MPI processes, more than the communication overhead. This is clearly visible in Figure~\ref{graph_mpi_time}.

\subsection{Energy considerations}

Knowing the importance that energy efficiency has in modern HPC systems, we also study the energy consumption of the three applications.
Since the \dibona cluster offered a small amount of \skylake nodes (called \dibonaintel) with equivalent energy measuring capabilities than the \dibonaarm nodes, our measurements allow for a fair comparison of both platforms.

As expected, our energy measurements have been strongly affected by the performance of the applications.
Therefore, the maturity of the software stack has a substantial impact on the energy consumption, as vendor-specific tools deliver not only better performance but also lower energy-to-solution.
In this sense, we consider the example of Alya as a clear, successful story for the energy efficiency and the scalability of \arm systems on a complex production code.
For instance, we showed that the same simulation could be carried out on \dibonaarm saving $30\%$ of the energy and running $10\%$ slower compared to \dibonaintel nodes.
The tests of \graph highlight another success story for the energy efficiency of \dibonaarm, which also takes advantage of the fact that its node configuration offers a power of 2 number of cores.
In general, we have observed that the energy-to-solution for the different applications is approximately the same on both platforms. However, if we consider the energy-delay-product, the better performance of \dibonaintel results also in a better efficiency, when compared to \dibonaarm.
%

\subsection{Lessons learned}

Considering the priorities of HPC-facility managers, we can conclude that \txtwo is a great leap forward for \arm architectures in this field. However, the energy efficiency is still behind the \skylake architecture delivered by Intel for pure HPC workloads. 

Thinking about the needs of domain scientists and HPC developers, the \txtwo processor is a platform worth considering as the maturity of the different software stacks is comparable to the state-of-the-art options available in Intel platforms. Programmability and stability of \dibonaarm nodes have been proved at the level of other production systems.

For computer architects, this study also offers some interesting insights. For instance, the importance of memory channels, as they can overcome the limitations of a slower clock and memory speeds. However, it also points out the demand for vector operations from current scientific applications, which means that improving the size of SIMD units will bring a significant advantage.

Globally, we think that the architectural point explored with the \txtwo CPU is extremely relevant for the research of a path towards Exascale. In our view, it shows that, for complex applications such as Alya, the performance penalties introduced by a smaller SIMD unit can be compensated by a higher memory bandwidth, and, more generally, it allows programmers and integrators to explore an innovative architectural design point that is able to deliver decent performance in a competitive power envelope.
We expect the situation will soon change when the new Scalable Vector Extension by \arm~\cite{rico2017arm} is implemented in some real hardware. However, we will reserve this discussion for future work.

\section*{Acknowledgments}
The authors thank the support team of Dibona operating at ATOS/Bull.
This work is partially supported by 
the Spanish Government through Programa Severo Ochoa (SEV-2015-0493), 
the Spanish Ministry of Science and Technology project (TIN2015-65316-P), 
the Generalitat de Catalunya (2017-SGR-1414),
the European Community's Seventh Framework Programme [FP7/2007-2013] and Horizon 2020 under the Mont-Blanc projects (grant agreements n.~288777, 610402 and 671697),
the European POP2 Center of Excellence (grant agreement n.~824080), and
the Human Brain Project SGA2 (grant agreement n.~785907).

\section*{References}

\bibliography{99-bibliography}

\end{document}